\documentclass[
showpacs,
pra,
twocolumn,
superscriptaddress,
aps,
amsmath,
amssymb
]{revtex4-2}

\usepackage[T1]{fontenc}
\usepackage[utf8]{inputenc}
\usepackage{newtxtext}
\usepackage{newtxmath}
\usepackage{microtype}

\usepackage{amsmath,amssymb,amsfonts}
\usepackage{mathtools}
\usepackage{bm}
\usepackage{physics}
\usepackage{bbold}

\usepackage{graphicx}
\graphicspath{{Figures/}}

\usepackage{float}
\usepackage{placeins}
\usepackage{multirow}
\usepackage{dcolumn}
\usepackage{booktabs}
\usepackage[caption=false]{subfig}
\usepackage{tikz}

\usepackage{xcolor}
\usepackage{pifont}
\usepackage{eurosym}

\usepackage[
breaklinks=true,
colorlinks=true,
citecolor=blue,
linkcolor=blue,
urlcolor=blue
]{hyperref}

\makeatletter
\renewcommand\@fnsymbol[1]{*}
\makeatother

\newcommand{\vect}[1]{\mathbf{#1}}

\DeclareMathAlphabet{\bi}{OML}{cmm}{b}{it}

\newcommand{\be}{\begin{equation}}
\newcommand{\ee}{\end{equation}}

\begin{document}
\title{Extended Haldane Model in The Dice Lattice: Multiple Flat-Band-Induced topological Transitions Revealed}
	\date{\today}
\author{Othmane Benhaida}
\email{othmane.benhaida@gmail.com}
\affiliation{LPHE, Modeling and Simulations, Faculty of Science, Mohammed V University in Rabat, Rabat, Morocco}
\affiliation{CPM, Centre of Physics and Mathematics, Faculty of Science, Mohammed V University in Rabat, Rabat, Morocco}
\author{Lalla Btissam Drissi}
\email[Corresponding author: ]{lalla-btissam.drissi@fsr.um5.ac.ma} % <-- l’email s’affichera sous le nom
\affiliation{LPHE, Modeling and Simulations, Faculty of Science, Mohammed V University in Rabat, Rabat, Morocco}
\affiliation{CPM, Centre of Physics and Mathematics, Faculty of Science, Mohammed V University in Rabat, Rabat, Morocco}
\affiliation{College of Physical and Chemical Sciences, Hassan II Academy of Sciences and Technology, Rabat, Morocco}
\author{El Hassan Saidi}
\affiliation{LPHE, Modeling and Simulations, Faculty of Science, Mohammed V University in Rabat, Rabat, Morocco}
\affiliation{CPM, Centre of Physics and Mathematics, Faculty of Science, Mohammed V University in Rabat, Rabat, Morocco}
\affiliation{College of Physical and Chemical Sciences, Hassan II Academy of Sciences and Technology, Rabat, Morocco}

	\begin{abstract}
In this study, we examine the introduction of the Haldane model into the dice lattice by altering the flow between the next-nearest-neighbour sites. This breaks the lattice's inversion and time-reversal symmetries. We demonstrate the presence of point-charge particle symmetries at $\phi^c=\pi/6$ and $5\pi/6$ and derive the analytical expression for quasi-energies. We demonstrate that a gap closure occurs at these critical points, inducing a topological transition. This is confirmed by calculating the Berry curvature and orbital magnetic moment. A topological analysis shows that the Chern numbers of the valence band $(\nu=0)$, the flat band $(\nu=1)$ and the conduction band $(\nu=2)$ depend strongly on the relationship between the fluxes $\phi^a $ and $\phi^c$. When $\phi^c = \phi^a$, the Chern numbers are $(C_0, C_1, C_2) = (2, -2, 0)$ in the region $\phi^c \in [0, \pi/6[$, and (0, 2, -2) in the region $\phi^c\in ]5\pi/6, \pi]$. Conversely, when $\phi^c \neq \phi^a$, the topological invariants become $ (C_1, C_2) = (-1, -1)$ for $\phi^c \in [0, \pi/6[$, and $(C_0, C_1, )= (1, 1)$ for $\phi^c\in ]5\pi/6, \pi]$. These variations reflect topological phase transitions at the critical points $\phi^c=\pi/6$ and $5\pi/6$, affecting all of the system's bands. Furthermore, the anomalous Hall conductivity exhibits a quantized plateau of 2$\sigma_{0}$, as well as an unquantized tilted plateau evolving from 1.50$\sigma_{0}$ to 1.25$\sigma_{0}$ at the same transition points. Controlling the flux allows topological transitions to be engineered and quantum transport in the dice lattice to be optimised, offering promising prospects for reconfigurable topological devices with low dissipation and robust quantum transport.\newline
		
	\end{abstract}
	\maketitle
\section{INTRODUCTION}
Exploring and understanding topological quantum transitions in materials is a fundamental area of research in condensed matter physics due to the variety of their physical properties and their potential applications in the future. In recent years, extensive research has been conducted to deepen our understanding of topological phases and anomalous Hall effects (whether quantum or light-induced) in various systems, such as topological insulators, Kagome lattices, $\alpha -T_{3}$ structures, and moiré materials. These studies have illuminated the topological origin of the anomalous quantum Hall effect and the physical mechanisms responsible for the various classes of quantum anomalous Hall (QAH) insulators and their potential applications \cite{In1}. Quantum topological systems displaying the quantum Hall effect have also been observed in ferromagnetic semimetals \cite{In2}, and the optical control of topological phases in deformed  $\alpha -T_{3}$ lattices has been investigated \cite{In56}. Understanding the mechanisms of electron transport in quantum Hall insulators has focused on two main areas: chiral interactions at edges, including interference between chiral edge channels in small-width samples \cite{In3}, and the possibility of a quantum transition between two different topological states in the absence of a magnetic field \cite{In4}. Other studies have focused on engineering topological transitions in multi-orbital and multilayer lattices \cite{In4A, In5, In6,In57,JaBen}. The extension of the concept of topology to encompass localized states at corners or edges has given rise to the emergence of higher-order topological phases \cite{ In7A1, In7A2}, which differ from traditional edge or surface states in that they provide protection from perturbations \cite{ In7A1, In31, In7A3}. Haldane was the first to theoretically predict the existence of an anomalous quantum Hall insulator, which corresponds to the quantum Hall effect that occurs in the absence of an external magnetic field on a two-dimensional honeycomb lattice \cite{ In7}. This phenomenon can be attributed to the breaking of time-reversal symmetry, which is induced by the introduction of complex chiral hopping energies between the nearest neighbouring sites in non-trivial topological insulators \cite{ In7, In8, In10}. These insulators are characterized by a Hall resistance of $h/(Ce^2)$ and the presence of spin-polarized, dissipation-free chiral edge channels. The number of these channels is determined by the Chern number $C$ \cite{ In11, In12}. Thanks to these properties, the anomalous quantum Hall effect is expected to play a key role in the development of low-power and spintronic technologies, as well as future applications in quantum computing and communications. Initially, experimental studies of these systems were limited to Chern number $C=1$ \cite{In8}. However, this restriction was overcome thanks to theoretical predictions of phases with high Chern numbers ($C \geqslant 2$) \cite{In14,In15}, leading to their subsequent realization and observation in experiments \cite{In4,In13,In16}. These phases, known as high-Chern-number insulators (HCIs), provide a broader framework for exploring topological phenomena and their potential applications.

% With theoretical and experimental advances, systems with high Chern numbers ($C \geqslant 2$) have been predicted and observed \cite{In14, In15, In4, In13, In16}. These systems are known as high-Chern insulators (HCIs), and they could lead to new discoveries in physics and various future applications.\\
\par Thanks to the introduction of magnetic impurities in the topological insulator (TI)/TI layers \cite{In4}, the recently created topological insulator features an adjustable Chern number ranging from 1 to 2. This phenomenon modifies the coupling of adjacent QAH $C=1$ layers, leading to a phase transition
 in the middle TI layer in the absence of an external magnetic field. This makes it possible to create robust chiral junctions for use in low-power electronic and spintronic devices. The plateau transition leads to the development of the chiral edge channel, a fundamental aspect of zero-field QAH insulator physics. These topological insulators can also appear in flat-band systems, where they play a pivotal role in the physics of various materials. Notably, they are found in twisted layers that are responsible for superconductivity, as well as in other two-dimensional systems \cite{In17, In20}. This phenomenon has also been observed in graphene layers twisted at a magic angle in the presence of an external electric field \cite{In21, In22}. Furthermore, flat-band materials can undergo phase transitions and exhibit a high Chern number. 
 This has been observed in twisted bilayer graphene and multi-layer graphene stacked on boron nitride when an isolated flat strip is exposed to an external electric field \cite{In22, In23}. Flat-band systems can exhibit counterintuitive topologies and support various exotic states, including superconductivity \cite{In26, In27}, the fractional anomalous quantum Hall effect \cite{In28, In29} and the anomalous quantum Hall effect \cite{In30, In31}. The flat band structure is not limited to twisted bilayer graphene; it also exists in octagonal lattices, where interaction with nearest neighbours or application of magnetic flux can lead to topological isolation or higher-order topological phases \cite{In32}. It is also found in kagome lattices \cite{In33, In34}, Lieb lattices \cite{In37}, $T_{3}$ lattices \cite{In38,In38D} and $\alpha-T_{3}$ lattices \cite{IIn2}. The $\alpha-T_{3}$ lattice has recently been the subject of extensive study due to its flat band structure \cite{In40,In44}, which is based on controlling the coefficient $\alpha$ within the range [0, 1] \cite{In38R}. This two-dimensional semi-metallic lattice is considered a Dirac–Weyl semimetal and continuously transitions between a honeycomb graphene lattice ($\alpha = 0$) and a dice lattice  $T_{3}$ ($\alpha = 1$) \cite{In38, In46}.  Theoretical studies and experimental proposals have indicated its feasibility in artificial heterogeneous structures, as well as in optical lattices \cite{In49, In50}. In particular, a study of its magnetic properties revealed a transition from $\sigma_{xy} = 2n(2n + 1)e^2/h$ to $\sigma_{xy} = 4ne^2/h$ \cite{ In51}. In the Haldane model, the lattice exhibits a high Chern number ($C = 2$) associated with the dispersion bands, while the flat band structure is characterised by $C = 0$, resulting in a two-channel quantum Hall effect \cite{ IIn2}, which is absent in graphene. Additionally, a topological phase transition has been observed, from a non-trivial phase ($\sigma_{xy}=2e^{2}/h$) to a trivial phase ($\sigma_{xy}=0$) in the semi-Dirac limit \cite{ReffChern4A}. The combination of asymmetric sublattice symmetry breaking and $\alpha$ parameter control enables the engineering and modulation of chiral interface states in the $\alpha-T_3$ lattice. This has significant implications for topological systems \cite{Nascimento2025}. Electron-phonon coupling is a fundamental parameter that can induce, suppress, or transform first- and second-order topological phases by controlling gap closures \cite{Islam2024,Bhattacharyya2024}. Furthermore, applying alternating electric and magnetic potentials to a pseudospin-1 system with Rashba coupling induces distinct phase transitions characterised by chirality inversion and the emergence of unconventional anti-chiral edge states \cite{Parui2025}. Additionally, electric and magnetic fields activate and modulate the flat band of the $\alpha-T_3$ lattice, generating quantum transport properties that differ from graphene's \cite{Li2022}. Coupling parameters and symmetry breaking provide fine control over the energy spectrum and gap, paving the way for advanced band engineering in two-dimensional materials \cite{Cunha2021}. Introducing third-neighbour interactions into the semi-Dirac Haldane model also reveals Chern insulators ($|C|=1, 2$) and trivial phases accompanied by chiral modes and anomalous Hall conductivity \cite{In52}. The topological properties of the Haldane and modified Haldane models applied to the $\alpha-T_{3}$ lattice have been systematically investigated. It has been demonstrated that they each exhibit distinct phase transitions and that combining them generates three topological phases: the Chern insulator ($|C|=1$), the higher Chern insulator (($|C|=2$), and the topological metal (($|C|=2$). The emergence and stability of these phases are controlled by the $\alpha$ parameter \cite{In53}. Spin-orbit coupling also induces a topological transition at $\alpha = 0.5$, characterised by an abrupt change in the valley Chern number and spin Chern number (from $C_{s}^{2} =1$ to $C_{s}^{2} =2$). This transition is corroborated by clear signatures in Berry curvature, orbital magnetic moment (OMM), and optical absorption \cite{In54}. The topological transitions in the $\alpha-T_{3}$ system are studied under off-resonant circularly polarised light. Apart from the quantum Hall insulating phase with helical states, which is anomalous, there are also spin-polarised topological metallic phases \cite{In55}. Furthermore, it has been demonstrated that an $\alpha-T_{3}$ lattice subjected to off-resonance circular polarisation undergoes a topological Floquet transition at $\alpha = 1/\sqrt{2}$, with the Chern number changing from $C=1$ for $\alpha < 1/\sqrt{2}$ to $C = 2$ for $\alpha > 1/\sqrt{2}$ \cite{ReffHam4}. When the $\alpha-T_{3}$ lattice deforms while remaining subject to such polarisation, the Hall conductivity takes the values $\sigma_{xy}=e^2/h$ for $\alpha=0$ and $\sigma_{xy}=e^2/h$ for $\alpha=1$, with a topological transition at $\alpha = 1/\sqrt{2}$. In the semi-Dirac limit, however, the system becomes topologically trivial \cite{In56}. A similar transition has also been observed at $\alpha = 1/\sqrt{2}$ in an $\alpha-T_{3}$ bilayer subjected to off-resonance circular polarisation \cite{In57}. Furthermore, in a cyclic $\alpha-T_{3}$ bilayer, the Haldane model can produce a Chern number as high as $C = 5$, with an anomalous Hall conductivity of up to $\sigma_(xy) = 6e^2/h$  \cite{ReffChern5}. These results highlight the physical importance of the dice lattice($\alpha=1$). The lattice's high Chern number, flat bands, and diversity of topological transitions make it an ideal platform for studying topological phenomena and their transitions. Based on these results, we propose a new, unexplored version of the Haldane model by adjusting the phase of the next-nearest-neighbour flux(NNN). This reveals topological transitions that are new compared to the standard model, enabling control over transitions from trivial to nontrivial topological bands. The remainder of this work is devoted to a systematic and detailed analysis of this mechanism and its implications.
\par In systems subjected to a uniform magnetic field, the quantization of the Hall conductivity is traditionally attributed to the formation of Landau levels. However, in 1988, Haldane demonstrated that the quantum Hall effect could occur in the absence of net magnetic flux thanks to the breaking of time-reversal symmetry by a zero-mean periodic magnetic flux \cite{In7}. In this model, formulated on a honeycomb lattice, an on-site potential breaks the sublattice symmetry. At the same time, a complex modulation of hopping amplitudes between nearest neighbours induces a local flux that is cancelled at the mesh scale. This configuration preserves the flux average but breaks temporal 
 	\begin{figure*}[!t]
	\centering
	\vfill
	\includegraphics[width=1\linewidth]{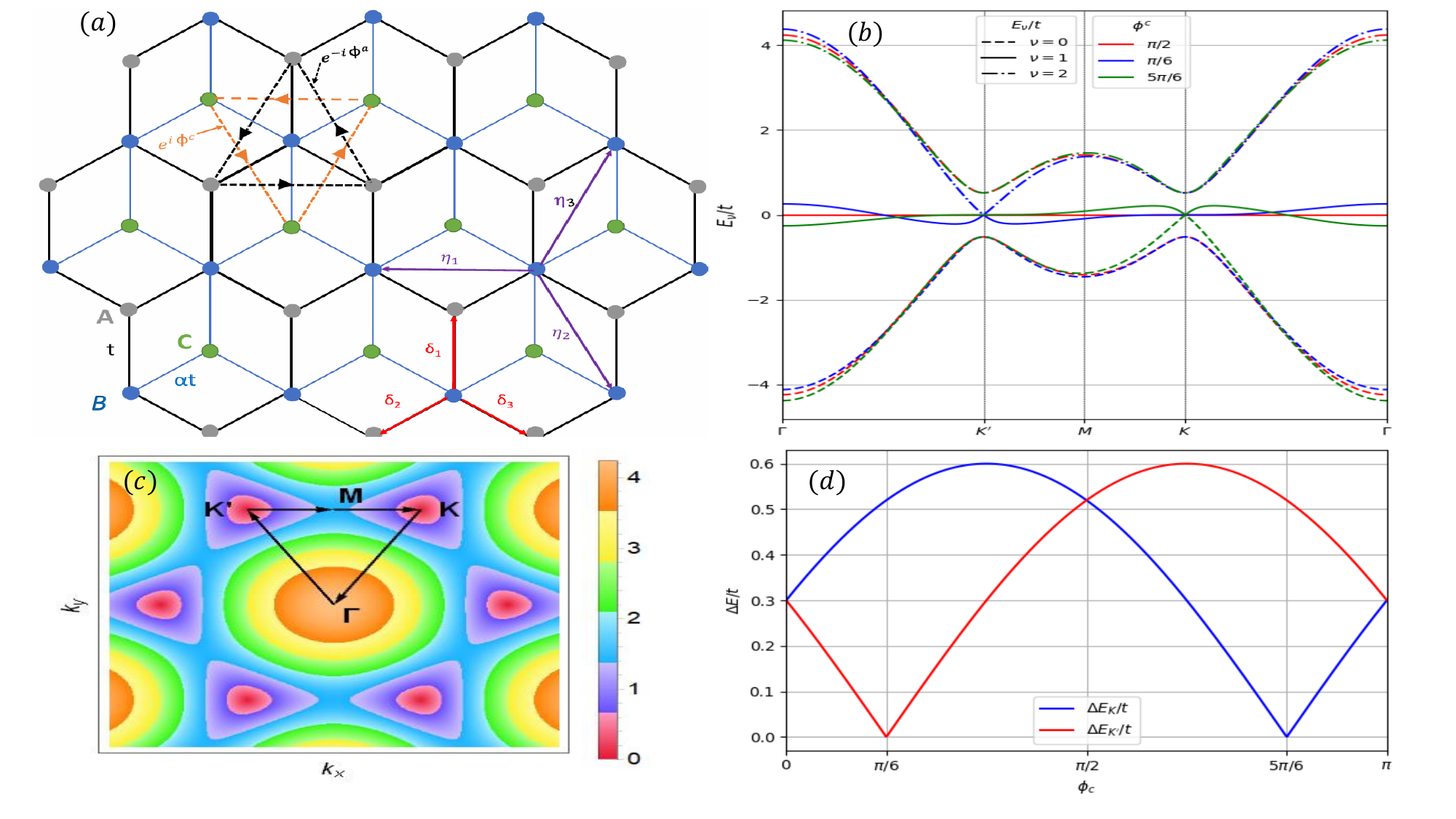}
	\vfill
	\caption{ (a) This is a schematic diagram of the $\alpha-T_{3}$ lattice, showing sublattices A, B, and C in grey, blue, and green, respectively. Nearest-neighbor (NN) hops between sites A and B are labelled t, and those between sites B and C are labelled $\alpha t$. The Haldane module introduces next-nearest neighbor hops : between sites A, hops are labelled $t_{a}e^{-i\phi^a}$ and represented by orange arrows, while between sites C, hops are labelled $t_{c}e^{i\phi^c}$ and represented by black arrows. The $\phi^a$ and $\phi^c$ phases are oriented counter-clockwise, as indicated by the arrows. (b) Shows the band structure of the $\alpha-T_{3}$ lattice, which consists of three bands: the valence band ($\nu =0$), the flat band ($\nu =1$), and the conduction band ($\nu =2$), represented by dashed, solid, and dash-dot lines, respectively. These bands are plotted for three flux values, $\phi^c: \pi/6, \pi/2$, and $5\pi/6$, which are represented by blue, red, and green lines, respectively. By setting the parameters to $\phi^a = \pi/2$, $\alpha = 1$, and $t_a = t_c = 0.1t$, the band structure is calculated along the$\bm\Gamma-\bm K'-\bm M-\bm K-\bm \Gamma$ high-symmetry path, as illustrated in subfigure (c). The density plot representation of the conduction band is also shown in the same subfigure, where each color level corresponds to a specific energy value, as indicated by the color bar (PlotLegends) in the sub-figure. (d) Plot showing the evolution of the two band gaps as a function of $\phi^c$. The first gap $\Delta E_{\bm K'}/t$, which is located at the Dirac point $\bm K'$, is shown in red. The second gap $\Delta E_{\bm K}/t$, which is located at the $\bm K$ point, is shown in blue. The parameters are the same in (b).}
	\label{Fig0}
\end{figure*}
and spatial inversion symmetries \cite{In7, IIn1}.
 In this study, we extend Haldane's model to an $\alpha-T_3$ lattice by modifying the flow associated with hops between NNN for sites A–A and C–C. This variation, combined with a significant potential difference between sites A and C, results in the simultaneous breaking of inversion symmetry and time reversal symmetry. We neglect the hops from the central site B to NNN because they are associated with three type A sites and three type C sites. The A–A and C–C hops generate a relatively high potential difference when attempting to hop from site B to a second neighbour. Due to this high energy barrier, the amplitude of the hop becomes very small and can be neglected in the model. This hypothesis has been adopted in previous studies \cite{IIn2,ReffChern4A,ReffChern5}. We studied two main cases: the symmetric case ($\phi^a = \phi^c$), in which the system reduces to the original Haldane model \cite{IIn2,ReffChern4A,ReffChern5}, and the asymmetric case ($\phi^a \neq \phi^c$). 
In the asymmetric case, the system exhibits quasi-metallic behaviour, characterised by contact between the energy bands at the Dirac points and a dispersed, flat band. This behaviour is absent from the original Haldane model when a potential is not applied to all sites. Our topological analysis is based on calculating the Berry curvature, orbital magnetic moment, Chern number, and anomalous Hall conductivity. These quantities are sensitive indicators of the system's topological nature. In the asymmetric case, the flat band exhibits a nonzero Berry curvature, inducing topological transitions at $\phi^c = \pi/6$ and $5\pi/6$, associated with a band inversion at the Dirac points. The orbital magnetic moment, which exhibits $C_3$-type rotational symmetry, also varies at these transition points, confirming the band inversion. The Chern number exhibits distinct variations at values $\phi^c = \pi/6$ and $\phi^c = \pi/6$, indicating a transition between trivial and non-trivial topological states depending on the values of $\phi^a$ and $\phi^c$. Finally, the anomalous Hall conductivity quantizes when the Fermi level is located in a band gap, exhibiting quantized plateaus. Conversely, when the Fermi level crosses a partially occupied band, the conductivity becomes unquantized. This control of flux opens up promising prospects for the development of low-dissipation reconfigurable electronic and optical devices, as well as for applications in the field of robust quantum transport.\\
The paper is structured as follows: Section \ref{secII} presents a theoretical and Hamiltonian model of the system. Section \ref{secIII} analyzes quasi-energy and band structure. Section \ref{secIV} examines the topological signatures: subsection \ref{sub-secI} deals with calculating the Berry curvature, and subsection \ref{sub-secII} with evaluating the orbital magnetic moment. Section \ref{secV} confirms phase transitions and explores the nature of topological states using the Chern phase diagram. Section \ref{secVI} studies anomalous Hall conductivity. Finally, Section \ref{secVII} provides a conclusion of the results.
\section{Theoretical model and Hamiltonian}\label{secII}
In this section, we propose a generalization of Haldane's model \cite{In7}. We assume that the hopping energy between second neighbors located on the same sublattice differs from that between second neighbors located on opposite sublattices. Similarly, the flux (or phase associated with the jump) is supposed to be distinct for each type of atomic site.
 We apply this model extension to the $\alpha-T_{3}$ lattice, which consists of three sub-lattices arranged in a hexagonal structure. Atoms located at the vertices of the hexagons are denoted A and B, while a third site, denoted C, is placed at the center of each hexagon. The hop energy between sites A and B is denoted by $t$, while that between B and C is given by $\alpha t$, where $\alpha$ is a coupling parameter. This parameter interpolates between two cases:  $\alpha=0$ corresponds to graphene and  $\alpha=1$ to the dice lattice. Fig.\ref{Fig0}(a) illustrates this structure. For a detailed description of this lattice, see reference \cite{In38R}. This lattice is characterized by three energy bands—two dispersive bands and one flat band—and provides an ideal platform for studying the extension of the Haldane model and analyzing its impact on the flat band.\newline

We define the extended tight-binding Hamiltonian of the Haldane model on the $\alpha-T_3$ lattice as follows
\begin{equation}
\begin{split}
H= &\,t\sum_{<i,j>} \left( a^{\dagger}_{i}b_{j}+\alpha c^{\dagger}_{i}b_{j}+h.c\right)+\\
&\sum_{<<i,j>>}\left( t_{a}e^{-i\phi^{a}}a^{\dagger}_{i}a_{j}+t_{c}e^{i\phi^{c}}c^{\dagger}_{i}c_{j}+h.c\right).
\end{split}
\end{equation}
where $d_{i}(d^{\dagger}_{i})$ represents the spinless fermionic annihilation (creation) operator at site $i$ belonging to sublattices $d = a, b, c$. The first term of the Hamiltonian corresponds to hopping to a nearest neighbour along directions $\bm\delta_{1}=a_{0}(0,1)$, $\bm\delta_{2}=a_{0}(-\sqrt{3}/2,-1/2)$, and $\bm\delta_{2}=a_{0}(\sqrt{3}/2,-1/2)$, where $a_{0}$ is the distance between two neighbouring sites.
Thus, the hopping amplitude between sublattices B and C in the $\alpha-T_3$ plane, in the -$\bm \delta_{1}$, -$\bm \delta_{2}$, and -$\bm \delta_{3}$ directions, is denoted $\alpha t$, while the hopping amplitude between sublattices A and B in the $\bm \delta_{1}$, $\bm \delta_{2}$, and $\bm \delta_{3}$ directions is denoted $t$. 
The second term of the Hamiltonian describes the complex hopping between the next-nearest neighbours and is an extension of the Haldane model. More precisely, $t_{a}$ is the hopping amplitude between sites A-A in the lattice under consideration, with the associated phase $\phi^{a}$, while $t_{c}$ is the hopping amplitude between sites C-C, with the associated phase $\phi^{c}$. The phase is considered positive for counterclockwise electron hopping, as illustrated in Fig.\ref{Fig0}(a). The NNN jump on sublattice B is neglected due to the high energy barriers surrounding it. The vectors connecting the nearest $A-A$ (or $C-C$) atoms are given by $\bm\eta_{1} = \bm\delta_{2}-\bm\delta_{3}$,  $\bm\eta_{2} = \bm\delta_{3}-\bm\delta_{1}$, and  $\bm\eta_{3} = \bm\delta_{1}-\bm\delta_{2}$.\\
\par The Fourier transform is utilised to express the diagonalised Hamiltonian in momentum space as $H=\sum_{k}\psi^{\dagger}(\bm k)\mathcal{H}(\bm k)\psi(\bm k)$, where $\psi(\bm k)=(a( \bm k),b(\bm k),c(\bm k))^{T}$ and $\mathcal{H}(\bm k)$ is the Bloch Hamiltonian, given as follows:
\begin{equation}\label{H1}
\mathcal{H}(\bm k,\phi^{a},\phi^{c})=\begin{pmatrix}
h_{z}(\bm k, t_a,\phi^{a}) & h(\bm k) &0\\
h^{*}(\bm k)& 0 &\alpha h(\bm k)\\
0 &\alpha h^{*}(\bm k)& h_{z}(\bm k, t_c,-\phi^{c})
\end{pmatrix}.
\end{equation}
where $h(\bm k)= h_{x}(\bm k)+ih_{y}(\bm k)$, the expressions $h_{x}(\bm k)$, $h_{y}(\bm k)$ and $h_{z}(\bm k, t_d,\phi^{d})$ are presented as follows:
\begin{align}
&h_{x}(\bm k)=2t\cos(\sqrt{3}k_{x}/2)\cos(k_{y}/2)+t\cos(k_{y}),\\
& h_{y}(\bm k)=2t\cos(\sqrt{3}k_{x}/2)\sin(k_{y}/2)-t\sin(k_{y}),
\end{align}
\begin{equation}
\begin{split}
h_{z}(\bm k, t_d,\phi^{d})=&2t_{d}\Bigr[ 2\cos{\dfrac{3k_y}{2}}\cos{(\dfrac{\sqrt{3}}{2}k_x+\phi^d)} +\\
&\cos{(\sqrt{3}k_x-\phi^d)}\Bigr].
\end{split} 
\end{equation}
where $h_{z}(\bm k, t_d,\phi^{d})$ represents $h_{z}(\bm k, t_a,\phi^{a})$ when $d=a$ and $h_{z}(\bm k, t_c,-\phi^{c})$) when $d=c$.\newline

\noindent Haldane proposed a model on graphene that exhibits quantized Hall conductivity, breaking the time-reversal symmetry via NNN hopping while zero net magnetic flux per unit cell is maintained. In this work, we aim to apply the Haldane model to the $\alpha-T_{3}$ lattice by modifying the NNN terms for sites belonging to different sublattices, i.e., by imposing $t_{a}\neq t_{c}$ and $\phi^a\neq \phi^c$, in this way investigating emergent topological phases. We also explore the influence of controllable parameters for visualizing topological phase transitions. To better understand the topological properties of the $\alpha-T_{3}$ lattice, we begin with an analysis of discrete symmetries for two distinct cases as follows:
\par
\noindent \textbf{First case}: we consider $t_{a}= t_{c}$ and $\phi^a= \phi^c=\pi/2$, a case previously studied in the context of the Haldane model. Here, we present the discrete symmetries of the $\alpha-T_{3}$ lattice in this context, fixing $\alpha=1$. These symmetries play a fundamental role in condensed matter physics, enabling electronic states to be classified. They include spatial inversion ($\mathcal{I}^{-1}\mathcal{H}(\bm k,\phi^{a},\phi^{c})\mathcal{I}=\mathcal{H}(-\bm k,\phi^{a},\phi^{c})$), time reversal ($\mathcal{T}^{-1}\mathcal{H}(\bm k,\phi^{a},\phi^{c})\mathcal{T}=\mathcal{H}(-\bm k,\phi^{a},\phi^{c})$), charge conjugation ($\mathcal{C}^{-1}\mathcal{H}(\bm k,\phi^{a},\phi^{c})\mathcal{C}=-\mathcal{H}(-\bm k,\phi^{a},\phi^{c})$), and the $C_{3 }$ rotational symmetry specific to the $T_{3}$ lattice. In this case, the time-reversal symmetry $\mathcal{T}=\mathcal{K}$—where $\mathcal{K}$ denotes the complex conjugation—is broken, while the inversion symmetry $\mathcal{I}$ and the charge conjugation $\mathcal{C}=\mathcal{U K}$ are preserved. We summarize these symmetries as follows \cite{ReffHamAdd1}:
\begin{equation}
\mathcal{I}=\begin{pmatrix}
0&0&1\\
0&1&0\\
1&0&0
\end{pmatrix}, \quad\mathcal{U}=\begin{pmatrix}
-1&0&0\\
0&1&0\\
0&0&-1
\end{pmatrix}.
\end{equation}
\textbf{Second case}: we consider $t_{a}= t_{c}$ and $\phi^a\neq \phi^c$, where $\phi^a$ is fixed at $\pi/2$ and $\phi^c$ takes the values $\phi^c_{1} = \pi/6$ and $\phi^c_{2} = 5\pi/6$. In studying the symmetries, we find that the inversion symmetry is broken, i.e., ($\mathcal{I}^{-1}\mathcal{H}(\bm k, \phi^a,\phi^c)\mathcal{I}\neq\mathcal{H}(-\bm k,\phi^a,\phi^c)$).
 Conversely, we identify a point-charge conjugation symmetry $( \mathcal{C}^{-1}\mathcal{H}(\bm k, \phi^a,\phi^c_{1})\mathcal{C}=-\mathcal{H}(-\bm k,\phi^a,\phi^c_{2}))$, as illustrated in Fig.\ref{Fig0}(a). This shows that the quasi-energy of the Hamiltonian is affected by this important symmetry. For each quasi-energy, there is a point of correspondence between the $\phi^c_{1}$ and $\phi^c_{2}$ configurations. This symmetry links the eigenvalues associated with $\phi^c_{1}$ to those of $\phi^c_{2}=\pi-\phi^c_{1}$.
Consequently, from this symmetry and equation (\ref{EQ12}), we can deduce the following relationships:
\begin{align}
&E_0(\bm k,\phi^a,\phi^c_1)=-E_2(-\bm k, \phi^a, \phi^c_2),\\
&E_1(\bm k, \phi^a, \phi^c_1)=-E_1(-\bm k, \phi^a, \phi^c_2).
\end{align}
From these symmetries, we can predict the existence of a topological phase transition. We intend to study this transition in the following sections.
\section{Quasi-energy  and band structure.}\label{secIII}
In this section, we analyze the behaviour of quasi-energies (i.e., eigenvalues) by fixing the parameter $\phi^a$ and varying the parameter $\phi^c$. To accomplish this, we diagonalize the Hamiltonian defined by equation (\ref{H1}) to determine the eigenvalues. This procedure yields the characteristic equation, a third-degree polynomial of the form $-E^3 + a_{2}E^2 + a_{1}E + a_{0} = 0$. We simplify it by introducing the substitution $E =\dfrac{a_{2}}{3} -\varepsilon$, which reduces it to the depressed cubic equation form $\varepsilon^{3} + p\varepsilon +q = 0$. Since the Hamiltonian is Hermitian, the solutions to this equation are real and can be expressed in trigonometric form:
\begin{equation}
\begin{split}
\varepsilon_{\nu}(\bm k, \phi^a,\phi^c)=&2\sqrt{-\dfrac{p(\bm k, \phi^a,\phi^c)}{3}}\cos\Bigr(-\dfrac{2\pi \nu}{3}+\\&\dfrac{1}{3}\arccos\Bigl(\dfrac{3q(\bm k, \phi^a,\phi^c)}{p(\bm k, \phi^a,\phi^c)}
\sqrt{\dfrac{-3}{p(\bm k, \phi^a,\phi^c)}}
\Bigr)\Bigr),
\end{split}
\end{equation}
with
\begin{equation}
a_2(\bm k, \phi^a,\phi^c)=h_z(\bm k,t_a,\phi^a)+h_z(\bm k, t_c,-\phi^c),
\end{equation}
\begin{equation}
a_1(\bm k, \phi^a,\phi^c)=-h_z(\bm k, t_a,\phi^a)h_z(t_c,-\phi^c)+|h(\bm k)|^{2}\left(1+\alpha^{2}\right),
\end{equation}
\begin{equation}
a_{0}(\bm k, \phi^a,\phi^c)=-|h(\bm k)|^{2}\left(\alpha^2 h_z(\bm k, t_a,\phi^a)+h_z(\bm k, t_c,-\phi^c)\right),
\end{equation}
\begin{equation}
p(\bm k, \phi^a,\phi^c)=-\left(a_1(\bm k, \phi^a,\phi^c)+\dfrac{a_{2}^{3}(\bm k, \phi^a,\phi^c)}{3}\right),
\end{equation}
\begin{equation}
\begin{split}
q(\bm k, \phi^a,\phi^c)=&a_0(\bm k, \phi^a,\phi^c)+\frac{a_1(\bm k, \phi^a,\phi^c)a_2(\bm k, \phi^a,\phi^c)}{3}+\\
&2\left(\dfrac{a_{2}(\bm k, \phi^a,\phi^c)}{3}\right)^{3}.
\end{split}
\end{equation}
Consequently, the eigenvalues of the Hamiltonian (\ref{H1}) and their associated normalized pseudo-eigenvectors are given by the following formula:
\begin{align}
\label{EQ12} E_{\nu}(\bm k, \phi^a,\phi^c)=\dfrac{a_{2}(\bm k, \phi^a,\phi^c)}{3}-\varepsilon_{\nu}(\bm k, \phi^a,\phi^c),
\end{align}
\begin{equation}
\begin{split}
\Psi_{\nu}(\bm k, \phi^a,\phi^c)=&\mathcal{N}_{\nu}(\bm k, \phi^a,\phi^c)\times\\
&\begin{pmatrix}
\dfrac{h(\bm k)}{E_{\nu}(\bm k, \phi^a,\phi^c)-h_z(\bm k, t_a,\phi^a)}\\ 1 \\\dfrac{\alpha h^{*}(\bm k)}{E_{\nu}(\bm k, \phi^a,\phi^c)-h_z(\bm k, t_c,-\phi^c)}
\end{pmatrix}.
\end{split}
\end{equation}
 where
 \begin{equation}
\begin{split}
\mathcal{N}_{\nu}(\bm k, \phi^a,\phi^c)=&\Bigr[ 1+\dfrac{|h(\bm k)|^{2}}{\left(E_{\nu}(\bm k, \phi^a,\phi^c)-h_z(\bm k, t_a,\phi^a)\right)^{2}}+\\
&\dfrac{\alpha^2 |h(\bm k)|^{2}}{\left(E_{\nu}(\bm k, \phi^a,\phi^c)-h_z(\bm k, t_c,-\phi^c)\right)^{2}}\Bigr]^{-1/2}.
\end{split}
\end{equation}
As for the limit $a_2(\bm k, \phi^a,\phi^c)/3$, from a physical perspective, this transformation does not alter the intrinsic properties of the system. The term $a_2(\bm k, \phi^a,\phi^c)/3$ acts as a global additive constant corresponding to a simple shift of the energy origin. Consequently, this does not affect either the relative band structure (gaps) or the topological properties.\newline
with $\nu = 0$, 1, and 2 corresponding, respectively, to the quasi-energies of the valence, flat, and conduction bands.\newline 
We analyze the band structure to identify the emergence of a topological transition in the first Brillouin zone, particularly in the low-energy regime. To this end, we set the parameters  $\alpha$ and $\phi^a$ to 1 and $\pi/2$, respectively, while varying parameter $\phi^c$ according to values $\pi/6$, $\pi/2$, and $5\pi/6$. We then calculate the quasi-energies representing the band structure in the first Brillouin zone by following the path connecting the high-symmetry points: $\bm\Gamma \left( 0, 0\right)$, $\bm K' \left( \dfrac{-2\pi}{3\sqrt{3}}, \dfrac{2\pi}{3}\right) $, $\bm M \left( 0,\dfrac{2\pi}{3}\right) $, and $\bm K \left( \dfrac{2\pi}{3\sqrt{3}}, \dfrac{2\pi}{3}\right) $, as exhibited in Fig.\ref{Fig0}(c). The quasi-energies are shown in Fig.\ref{Fig0}(b). Before introducing the extended Haldane term, we will now discuss the band structure, which exhibits semi-metallic behaviour characterized by the intersection of bands leading to degeneracy in the low-energy region, particularly at the $\bm K$ and $\bm K'$ points. The addition of the Haldane term breaks the time-reversal symmetry and opens an energy gap, thus lifting this triple degeneracy—particularly in the standard case where $\phi^c= \pi/2$.  However, for $\pi/6$ and $ 5\pi/6$, a different behaviour is observed: the flat band becomes dispersive and shows contact with the valence band at the $\bm K$ point for  $\phi^c= \pi/6$ and with the conduction band at the $\bm K'$ point for  $\phi^c= 5\pi/6$. This may indicate either a topological phase transition or a topological sign. The change in $\phi^c$ distorts the nature of the flat band, particularly at $\phi^c= \pi/6$  and $5\pi/6$. Additionally, when an off-resonance polarised light field is applied, this flat band contacts the conduction band at the Dirac $K'$ point and the valence band at the K point for $\alpha =1/\sqrt{2}$, confirming the occurrence of a topological transition \cite{ReffHam3,ReffHam4,In56,In57}.  In this model, consequently, inversion symmetry is preserved for $\phi^c= \pi/2$, implying $E_\nu(\bm k, \phi^a,\phi^c)=E_\nu(-\bm k, \phi^a,\phi^c)$, as illustrated in Fig.\ref{Fig0}(b). However, it is broken for $\phi^c_{1} = \pi/6$ and $\phi^c_{2} = 5\pi/6$. Furthermore, point-charge conjugation symmetry is observed via the relations  $E_0(\bm k, \phi^a,\phi^c_{1})=-E_2(-\bm k, \phi^a,\phi^c_{2})$ and $E_1(\bm k, \phi^a,\phi^c_{1})=-E_1(-\bm k, \phi^a,\phi^c_{2})$.\newline

\par To determine whether a topological phase transition is occurring, as illustrated in Fig.\ref{Fig0}(b), we examine two distinct gaps situated close to the Dirac points $\bm K'$
 and $\bm K$, corresponding to the values $\phi^{c}_1=\pi/6$ and $\phi^{c}_2=5\pi/6$, respectively. 
 These gaps are evaluated based on the function of the parameter $\phi^c$, which makes it possible to analyze a phase transition characterized by the closing of one of the gaps. The first gap, located at the $\bm K'$ point, is defined as the difference between the conduction band ($E_2(\bm K',\phi^a,\phi^c)$) and the flat band ($E_1(\bm K',\phi^a,\phi^c)$) near the Dirac $\bm K'$ point. This difference, denoted as $\Delta E_{\bm K'} = E_2(\bm K',\phi^a,\phi^c)-E_1(\bm K',\phi^a,\phi^c)$, is calculated numerically and exhibits behavior similar to that of $|3t_c\left(\cos\phi^c-\sqrt{3}\sin\phi^c\right)|$. This gap becomes zero when $\phi^{c}_1=\dfrac{\pi}{6}$, as shown in Fig.\ref{Fig0}(d). It decreases quasi-linear (or quasi-sinusoidal behaviour)  in the interval $0<\phi^c<\dfrac{\pi}{6}$, exhibiting sinusoidal behaviour beyond this interval. The second gap, located near the Dirac point $\bm K$, is defined as the difference between the flat band $E_1(\bm K,\phi^a,\phi^c)$ and the valence band $E_0(\bm K,\phi^a,\phi^c)$). This gap exhibits analogous behaviour $|3t_c\left(\cos\phi^c+\sqrt{3}\sin\phi_c\right)|$. The second gap is given by $\Delta E_{\bm K} = E_0(\bm K,\phi^a,\phi^c)-E_1(\bm K,\phi^a,\phi^c)$. This second gap becomes null for $\phi^{c}_2=\dfrac{5\pi}{6}$, exhibits sinusoidal behaviour in the interval $0<\phi^c<\dfrac{5\pi}{6}$, and increases quasi-linear in the subsequent interval.\newline

Based on the results in  Fig.\ref{Fig0}(b)) and \ref{Fig0}(d), a transition in the band topology is possible when the gap opens and closes at points $\bm K$ and $\bm K'$. The gap opens when parameter $\phi^{c}$ varies continuously except for two specific values: $\phi^{c}_{1}$ and $\phi^{c}_{2}$. At these values, the gap closes. Therefore, we expect a phase transition at $\phi^{c}_{1}$ and $\phi^{c}_{2}$. To investigate this, we will study the Berry curvature and the magnetic orbital moment to determine if a phase transition or topological properties are present. This will allow us to confirm or refute the existence of a topological transition using a topological invariant: the Chern number.
\section{Topological signatures}\label{secIV}
The aim of this section is to study the topological features of the $\alpha-T_3$ system, with a particular emphasis on the behaviour of the Berry curvature and the orbital magnetic moment. These features are essential for detecting and identifying topological phases.
\subsection{Berry curvature}\label{sub-secI}
\begin{figure*}[!t]
	\centering
	\includegraphics[width=0.7\linewidth]{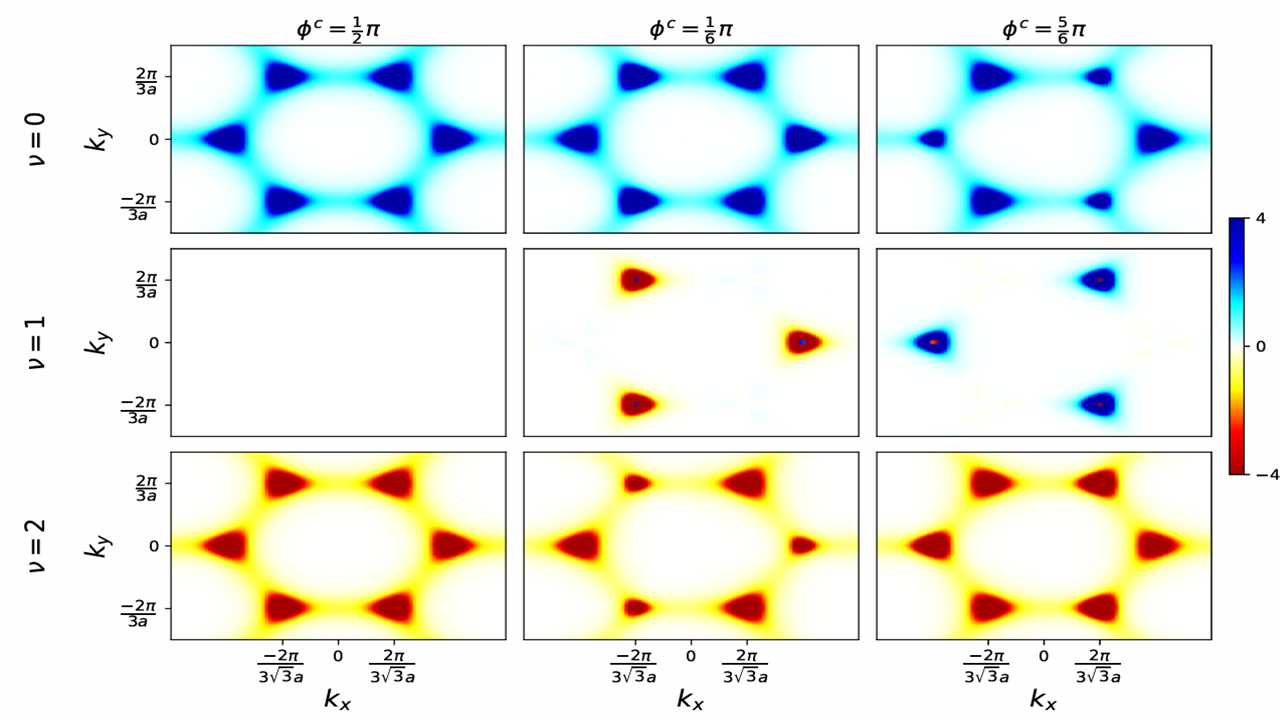}
	\caption{A density plot of the Berry curvature in $k_{x}-k_{y}$-space is shown, corresponding to the valence band ($\nu = 0$), the flat band ($\nu=1$), and the conduction band ($\nu=2$). The results are presented for three cases: $\phi^{c}=\pi/2$ (the normal case, where $\phi^{c} =\phi^{a}$), $\phi^{c}=\pi/6$ (the first topological transition, where $\phi^{c}\neq\phi^{a}$), and $\phi^{c}=5\pi/6$ (the second topological transition, where $\phi^{c} \neq\phi^{a}$). The other parameters are fixed at $\alpha=1$, $\phi^{a}=\pi/2$, and $t_{a} = t_{c} = 0.1t$.}
	\label{Fig3}
\end{figure*}

Berry curvature constitutes a fundamental notion in topological quantum matter physics. It is a geometric property that describes the behaviour of uncorrelated electrons and emerges from the Berry phase. Electrons acquire the Berry phase as they move through a periodic potential or crystal lattice. More precisely, the Berry phase corresponds to the phase accumulated by the Bloch wavefunction along a closed path in $\bm k$-space, within the first Brillouin zone, under adiabatic conditions \cite{ReffBerry1}. 
The Berry curvature $\Omega_{\nu}(\bm k, \phi^a,\phi^c)$, expressed as $i\nabla_{\bm k}\times\bra{\Psi_{\nu}(\bm k, \phi^a,\phi^c)}\nabla_{\bm k}\ket{\Psi_{\nu}(\bm k, \phi^a,\phi^c)}$
 , is based on the periodic part of the Bloch function $\Psi_{\nu}(\bm k, \phi^a,\phi^c)$ and is invariant under  gauge-transformations.
 Its function is comparable to that of an effective magnetic field regarding the movement of electrons in real space. This concept is crucial in several physical phenomena, including the anomalous Hall effect \cite{In6}, spin transport \cite{ReffBerry3}, and spin-orbit coupling \cite{ReffBerry4}, which give rise to significant topological effects. In two-dimensional systems, the Berry curvature is generally transverse and oriented along the $\bm z$ component. It can be calculated numerically using the following expression \cite{ReffBerry1}:
\begin{equation}\label{EQ16}
\Omega_{\nu}(\bm{k},\phi^a,\phi^c)
=
-2\hbar^2 \,\mathrm{Im}
\sum_{\nu'\neq\nu}
\frac{
	M^{x}_{\nu\nu'}M^{y}_{\nu'\nu}
}{
	(E_{\nu}(\bm k, \phi^{a,c})-E_{\nu^{\prime }}(\bm k, \phi^{a,c}))^{2}
},
\end{equation}
with
\begin{equation}
M^{i}_{\nu\nu'}
=
\bra{\Psi_\nu(\bm{k},\phi^a,\phi^c)}
v_i
\ket{\Psi_{\nu'}(\bm{k},\phi^a,\phi^c)}.
\end{equation}
where $v_{i}$ is the in-plane electron velocity in the direction $i=x,y$, is given by $v_{x}\hbar^{-1}\nabla_{k_{i}}\mathcal{H}(\bm k,\phi^{a},\phi^{c})$. Berry curvature describes the geometrical response of electrons, i.e., how the electron wavefunction reacts to band curvature when the wave vector $\bm k$ in a material is varied. It also provides essential information on phase transitions, particularly when bands become degenerate, resulting in singularities in the first Brillouin zone. This behavior is fundamental to understanding the system's topological properties. Additionally, analyzing fundamental properties requires us to consider the system's symmetry constraints.  The symmetries considered here are spatial inversion $\mathcal{I}$, time reversal $\mathcal{T}$, and charge conjugation $\mathcal{C}$, whose actions on the Berry curvature are written respectively as:
$\mathcal{I}^{-1}\Omega_{\nu}(\bm  k,\phi^{a},\phi^{c})\mathcal{I}= \Omega_{\nu}(-\bm  k,\phi^{a},\phi^{c})$, $\mathcal{T}^{-1}\Omega_{\nu}(-\bm k,\phi^{a},\phi^{c})\mathcal{T}= -\Omega_{\nu}(-\bm  k,\phi^{a},\phi^{c})$, and $\mathcal{C}^{-1}\Omega_{\bar \nu}(\bm k,\phi^{a},\phi^{c})\mathcal{C}=-\Omega_{\nu}(-\bm  k,\phi^{a},\phi^{c})$, where $\bar{\nu}$ is associated with the conjugate energy band $-E_\nu(-k,\phi_a,\phi_c)$. When the symmetries $\mathcal{T}$ and $\mathcal{I}$ are simultaneously preserved, the Berry curvature vanishes over the entire first Brillouin zone. In fact, $\Omega_\nu$ is an odd function under the operation $\mathcal{T}$ and an even function under the operation $\mathcal{I}$. Therefore, under the combined symmetry $\mathcal{TI}$, the relation becomes $\Omega_\nu(k,\phi_a,\phi_c) = -\Omega_\nu(k,\phi_a,\phi_c)$, a constraint that necessarily imposes $\Omega_\nu(k,\phi_a,\phi_c) = 0$. On the other hand, breaking one of these symmetries (notably $\mathcal{T}$) allows the emergence of a nonzero Berry curvature. This condition is necessary for the appearance of a topological phase, because the integration of $\Omega_\nu(k,\phi_a,\phi_c)$ over the Brillouin zone can then result in a non-trivial Chern number.
When considering the generalized Haldane term applied to the $\alpha-T_{3}$ lattice, the parameters are set to $\alpha =1$, $\phi^a= \pi/2$ and $\phi^c=\pi/6$, $\pi/2$ and $5\pi/6$. In this study, we examine the behaviour of the Berry curvature in the standard Haldane model case, where $\phi^c = \phi^c =\pi/2$, as illustrated in Fig.\ref{Fig3}. In this case, the Berry curvature is non-zero for both the conduction band ($\nu = 2$) and the valence band ($\nu = 0$), resulting in symmetry breaking, particularly about time-reversal symmetry. In this context, the Berry curvature is concentrated around the Dirac points distributed in momentum space and behaves like a monopole. We also observe that inversion symmetry ($\mathcal{I}^{-1}\Omega_{\nu}(\bm  k,\phi^{a},\phi^{c})\mathcal{I}= \Omega_{\nu}(-\bm  k,\phi^{a},\phi^{c})$ ) is preserved, as are conjugate charge ($\mathcal{C}^{-1}\Omega_{2}(\bm  k,\phi^{a},\phi^{c})\mathcal{C}= -\Omega_{0}(-\bm  k,\phi^{a},\phi^{c})$, $\mathcal{C}^{-1}\Omega_{1}(\bm  k,\phi^{a},\phi^{c})\mathcal{C}= -\Omega_{1}(-\bm  k,\phi^{a},\phi^{c})$) and rotation ($C_{3}$) symmetries.\newline
In the case of the generalised Haldane model, when $\phi^{c}\neq\phi^{a}$, we set  $\phi^{a}=\pi/2$ and vary $\phi^{c}$ between two values: $\phi_{1}^{c}=\pi/6$ and $\phi_{2}^{c}=5\pi/6$. As in the previous case, we study the Berry curvature, as shown in Fig.\ref{Fig3}.
The Berry curvature is non-zero and concentrated at the Dirac points $\bm K$ and $\bm K'$ for the valence band ($\nu=0$) and the conduction band ($\nu=2$). However, unlike in the case of $\phi^{c}=\phi^{a}=\pi/2$, the flat band ($\nu = 1$) also exhibits non-zero Berry curvature, which was not previously observed. This behaviour results from breaking the inversion symmetry ($\mathcal{I}^{-1}\Omega_{\nu}(\bm k,\phi^{a},\phi^{c})\mathcal{I}\neq \Omega_{\nu}(-\bm  k,\phi^{a},\phi^{c})$ ). Non-zero Berry curvature leads to the emergence of conducting surface states, which are protected by the topology of the band structure. Additionally, point-charge symmetries ($\mathcal{C}^{-1}\Omega_{2}(\bm  k,\phi^{a},\phi^{c}_{1})\mathcal{C}= -\Omega_{0}(-\bm  k,\phi^{a},\phi^{c}_{2})$, $\mathcal{C}^{-1}\Omega_{1}(\bm  k,\phi^{a},\phi^{c}_{1})\mathcal{C}= -\Omega_{1}(-\bm  k,\phi^{a},\phi^{c}_{2})$) and rotation symmetry ($C_{3}$) are observed to be preserved. Breaking the inversion symmetry changes the topology of the flat band from a trivial state for $\phi^{c}=\pi/2$ to a non-trivial state for $\phi^{c}_{1}=\pi/6$ and  $\phi^{c}_{2}=5\pi/6$. This change is accompanied by a reversal in the sign of the Berry curvature, which is negative for $\phi^{c}_{1}=\pi/6$ and positive for $\phi^{c}_{2}=5\pi/6$. This indicates a topological phase transition linked to a band inversion affecting the flat band.\newline
\par
\begin{figure}[t]
	\centering
	\includegraphics[width=1\linewidth]{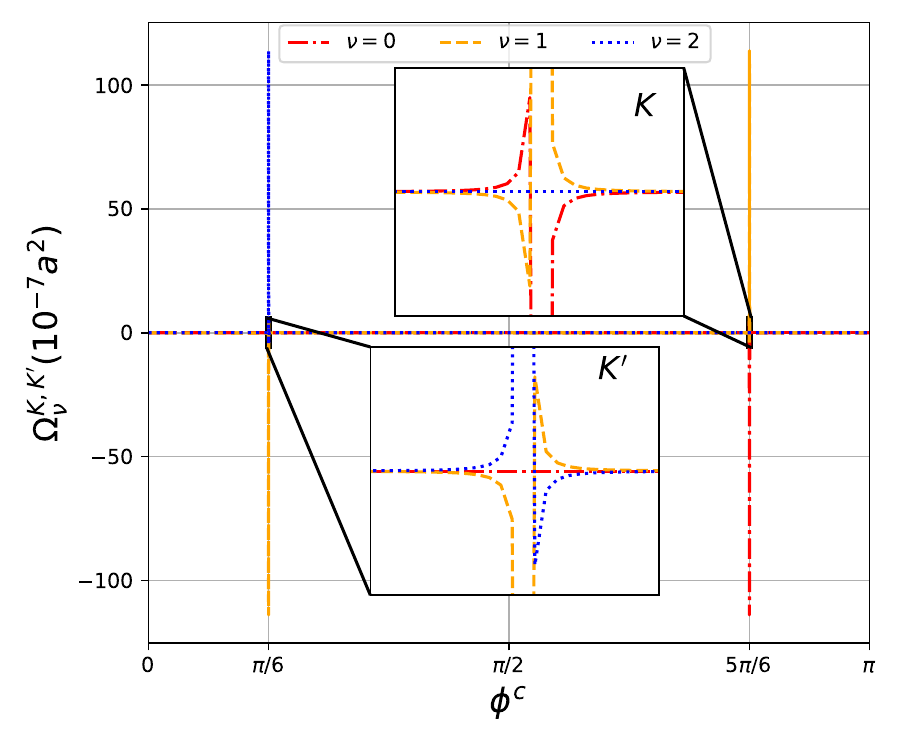}
	\caption{The evolution of Berry curvature as a function of $\phi^{c}$ at Dirac points $\bm K$ and $\bm K'$ is represented, as shown in the zoom. The Berry curvature is calculated for all three bands: the valence band ($\nu = 0$), the flat band ($\nu = 1$), and the conduction band ($\nu = 2$). These are represented by dash-dot, dashed, and dotted lines, respectively. Two topological phase transitions are observed: the first at $\phi^{c}=\pi/6$ at the $\bm K'$ point and the second at $\phi^{c}=5\pi/6$, also at the $\bm K$ point. The other parameters are identical to those in Fig.\ref{Fig3}.}
	\label{Fig33}
\end{figure}

 To illustrate this phase transition, we plot the Berry curvature as a function of $\phi^{c}$. This is essentially manifested in the vicinity of the Dirac points $\bm K$ and $\bm K'$. Fig.\ref{Fig33} shows how the Berry curvature evolves as $\phi^{c}$ varies continuously. Two sharp peaks are observed: the first at $\phi^{c}_{1}=\pi/6$ near the $\bm K'$ point and the second at $\phi^{c}_{2}=5\pi/6$ near the $\bm K$ point.
At the $\bm K'$ point ($\phi^{c}_{1}=\pi/6$), the Berry curvatures associated with the flat and conduction bands diverge, while that of the valence band remains finite. Conversely, at the $\bm K$ point ($\phi^{c}_{2}=5\pi/6$), it is the Berry curvatures of the flat and valence bands that diverge while that of the conduction band remains finite. These divergences occur precisely when the bands touch, coinciding with the closure of the gaps at $\phi^{c}_{1}=\pi/6$ and $\phi^{c}_{2}=5\pi/6$ observed in Fig.\ref{Fig0}(d).
Furthermore, the change in sign of the Berry curvature can be seen: near the $\bm K'$ point, the curvature becomes positive (negative) for the conduction (flat) band when $\phi_{c}^{1}$, and then becomes negative (positive). Near $\bm K$, we observe symmetrical behaviour, where the Berry curvature of the valence (flat) band changes from a positive (negative) to a negative (positive) sign, and vice versa, as it crosses $\phi^{c}_{2}=5\pi/6$. These sign inversions are confirmed by the observed divergences. They indicate the existence of a topological phase transition. This transition takes place at points $\phi_{1}^{c}$ and $\phi_{2}^{c}$.
\subsection{Orbital Magnetic Moment}\label{sub-secII}
\begin{figure*}[!t]
	\centering
	\includegraphics[width=0.7\linewidth]{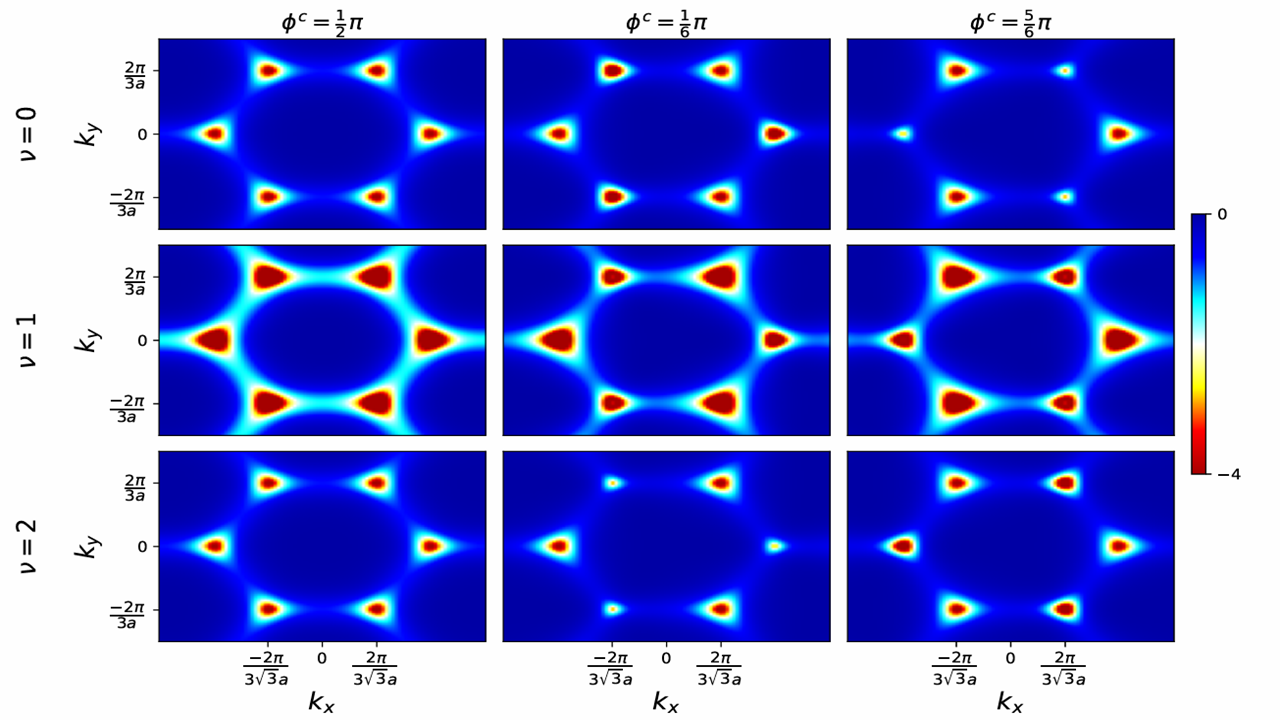}
	\caption{ The density plot of the magnetic orbital moment (OMM) in $k_{x}-k_{y}$ space is shown. This corresponds to the valence band ($\nu = 0$), the flat band ($\nu = 1$), and the conduction band ($\nu = 2$). The cases considered are: $\phi^c=\pi/2$ (normal case, where $\phi^a = \phi^c$ and inversion symmetry is preserved); $\phi^c=\pi/6$ (first transition, where $\phi^a \neq \phi^c$ and inversion symmetry is broken);  $\phi^c=5\pi/6$ (second transition, where inversion symmetry is broken). The other parameters are fixed at $\alpha=1$,  $\phi^a=\pi/2$, and $t_{a}= t_{c}=0.1t$.}
	\label{Fig4}
\end{figure*}

The orbital magnetic moment is an intrinsic property of Bloch bands and is represented by a wave packet in the semi-classical framework. This is the result of the autorotation of this wave packet around its center of mass \cite{ReffOMR1}. Bloch electrons have been shown to carry an orbital magnetic moment, expressed as $\mathcal{M}_{\nu}(\bm  k,\phi^{a},\phi^{c})=-\dfrac{i e}{2\hbar} \left\langle  \nabla_{\bm k} \Psi_{\nu}(\bm k) \middle| \times \left( \mathcal{H}(\mathbf{k}, \phi^{a}, \phi^{c}) - E_{\nu} \right) \middle| \nabla_{\bm k} \Psi_{\nu}(\bm k) \right\rangle$. 
The orbital magnetic moment is closely related to the geometry of the electronic band structure and the Berry curvature. This implies that $\mathcal{M}_{\nu}(\bm  k,\phi^{a},\phi^{c}) \propto\Omega_{\nu}(\bm  k,\phi^{a},\phi^{c})E(\bm k)$\cite{ReffOMR2} . For two-dimensional Dirac systems, Xiao\cite{Xiao2007}. showed that the orbital magnetic moment associated with each valley is directly related to the Berry curvature via a proportionality relation involving the band energy. $\mathcal{M}_{\nu}(\bm  k,\phi^{a},\phi^{c}) =-\dfrac{e}{\hbar}\Omega_{\nu}(\bm  k,\phi^{a},\phi^{c})E_{\nu}(\bm k)$, in the case of two-band effective models (e.g., graphene).  Similarly,  Chang \cite{ReffOMR1} showed that the magnetic moment can be derived from the Berry curvature via interband energy differences ($E_{\nu'}(\bm k)-E_{\nu}(\bm k))$. The orbital magnetic moment is important for various phenomena, such as the normal Hall effect \cite{ReffOMR3} and chiral magnetic interactions \cite{ReffOMR4}. In a two-dimensional system, this moment is perpendicular to the plane, i.e. it is oriented along the $z$-axis. It can then be written as \cite{In57}:
\begin{equation}  
\mathcal{M}_{\nu}(\bm  k,\phi^{a},\phi^{c})=-\hbar e\mathfrak{Im}\sum_{\nu^{\prime }\neq \nu}\frac{
	M^{x}_{\nu\nu'}M^{y}_{\nu'\nu}
}{
	E_{\nu}(\bm k, \phi^{a,c})-E_{\nu^{\prime }}(\bm k, \phi^{a,c})
}.
\end{equation}
We will now examine how the orbital magnetic moment (OMM) behaves in the extended Haldane model when applied to the $\alpha-T_{3}$ lattice. First, we set $\alpha = 1$ and $\phi^a = \pi/2$, and then vary $\phi^c$ according to three values: $\pi/6$, $\pi/2$, and $5\pi/6$. The orbital magnetic moment provides information that is complementary to that given by the Berry curvature. The OMM is subject to certain symmetry constraints, being invariant under inversion  ($\mathcal{I}^{-1}\mathcal{M}_{\nu}(\bm  k,\phi^{a},\phi^{c})\mathcal{I}= \mathcal{M}_{\nu}(-\bm  k,\phi^{a},\phi^{c})$), charge conjugation symmetry  ($\mathcal{C}^{-1}\mathcal{M}_{\bar\nu}(\bm  k,\phi^{a},\phi^{c})\mathcal{C}= \mathcal{M}_{\nu}(-\bm  k,\phi^{a},\phi^{c})$), and time reversal symmetry  ($\mathcal{T}^{-1}\mathcal{M}_{\nu}(\bm  k,\phi^{a},\phi^{c})\mathcal{T}= -\mathcal{M}_{\nu}(\bm  k,\phi^{a},\phi^{c})$). Fig.\ref{Fig4} shows the distribution of the OMM in $k_{x}-k_{y}$ space. We observe time-reversal symmetry breaking, which manifests as a non-zero OMM localized mainly at the $\bm K$ and $\bm K'$ Dirac points.
In the standard case where $\phi^a = \phi^c=\pi/2$, the OMM is associated with the conduction and valence bands. Their behaviour coincides, indicating that both inversion symmetry  ($\mathcal{I}^{-1}\mathcal{M}_{0,2}(\bm  k,\phi^{a},\phi^{c})\mathcal{I}= \mathcal{M}_{0,2}(-\bm  k,\phi^{a},\phi^{c})$) and charge conjugation symmetry  ($\mathcal{C}^{-1}\mathcal{M}_{0(2)}(\bm  k,\phi^{a},\phi^{c})\mathcal{C}= \mathcal{M}_{2(0)}(-\bm  k,\phi^{a},\phi^{c})$) are preserved. There is also a non-zero OMM associated with the flat band. However, this phenomenon is not apparent in the Berry curvature, even though the inversion($\mathcal{I}^{-1}\mathcal{M}_{1}(\bm  k,\phi^{a},\phi^{c})\mathcal{I}= \mathcal{M}_{1}(-\bm  k,\phi^{a},\phi^{c})$), charge conjugation($\mathcal{C}^{-1}\mathcal{M}_{1}(\bm  k,\phi^{a},\phi^{c})\mathcal{C}= \mathcal{M}_{1}(-\bm  k,\phi^{a},\phi^{c})$), and $C_{3}$ symmetries are also respected. We know that the Berry curvature measures the geometric torsion of the wave function in the eigenvector space. The symmetry of the system can result in the local cancellation of curvature, as in the case of inversion symmetry combined with time reversal. Thus, the global topology can be zero ($\Omega_{0,1,2}(\bm k, \phi^a,\phi^c)=0$). However, the orbital magnetic moment,$\mathcal{M}_{0}(\bm  k,\phi^{a},\phi^{c})$, arises from the self-rotating circulation of the wave packet around its centre. This circulation describes an internal, microscopic current associated with the Bloch state, which generally has an asymmetric internal structure that generates local circulation. Consequently, even when the Berry curvature is zero ($\Omega_{0}(\bm k, \phi^a,\phi^c)=0$), the orbital magnetic moment can remain nonzero because the electron can still exhibit intrinsic local rotation.\newline
When $\phi^{a} \neq \phi^{c}$, we set $\phi^{a} = \pi/2$ , $\phi^{c}_{1} = \pi/6$ and $\phi^{c}_{2}= 5\pi/6$. We then observe that the OMM value differs for each corresponding band, indicating a breaking of the inversion symmetry. However, we identify a conjugate point-charge symmetry, represented by $\mathcal{C}^{-1}\mathcal{M}_{2}(\bm  k,\phi^{a},\phi_{1}^{c})\mathcal{C}= \mathcal{M}_{0}(-\bm  k,\phi^{a},\phi_{2}^{c})$ and $\mathcal{C}^{-1}\mathcal{M}_{1}(\bm  k,\phi^{a},\phi_{1}^{c})\mathcal{C}= \mathcal{M}_{1}(-\bm  k,\phi^{a},\phi_{2}^{c})$. Breaking the inversion symmetry induces a change in the OMM values associated with the conduction and flat bands at the Dirac point $\bm K'$ for $\phi^{c}= \pi/6$, as well as in the valence and flat bands at the $\bm K$ point for $\phi^{c}= 5\pi/6$. These values were indeed not chosen arbitrarily. They are related by the transformation $\phi_{2}^{c}=\pi-\phi_{1}^{c}$, which preserves charge conjugate symmetry while breaking inversion symmetry. These values correspond to critical points  
 at which the gap closes (see Fig.\ref{Fig0}(d)), a phenomenon that can indicate a phase transition or band inversion associated with 
  \begin{figure}[H]
 	\centering
 	\includegraphics[width=1\linewidth]{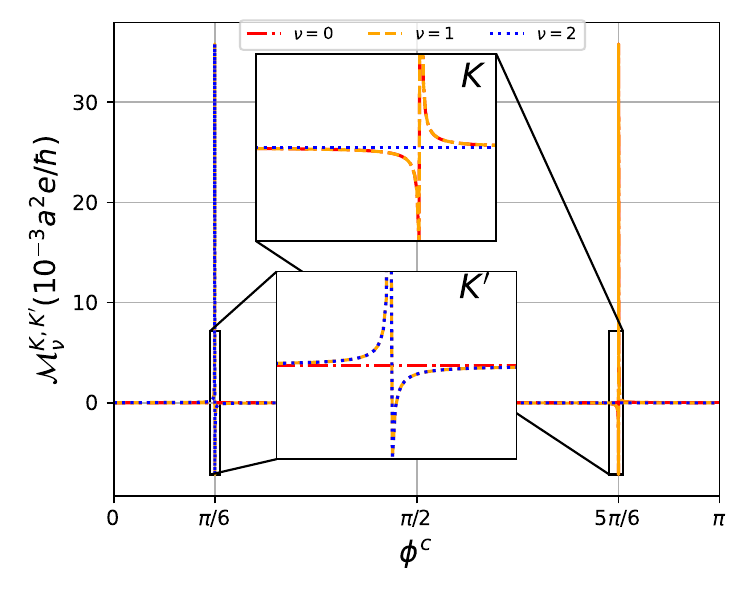}
 	\caption{(b) The magnetic orbital moment's evolution as a function of $\phi^c$ is shown in the zoom, at the Dirac points $\bm K$ and $\bm K'$. The OMM is calculated for the valence band ($\nu = 0$), the flat band ($\nu = 1$) and the conduction band ($\nu = 2$) and is represented by dashed, dotted and mixed dotted lines, respectively. The other parameters are identical to those in Fig.\ref{Fig4}.}
 	\label{Fig44}
 \end{figure}
 \noindent the breaking of inversion symmetry. 
 Therefore, the appearance  of these values stems directly from the symmetry properties of the system—specifically, point symmetry and the breaking of inversion symmetry—which lead to the closing of the gap at these critical points. They therefore result from the intrinsic structure of the model and not from an arbitrary choice. To determine the possible existence of a topological phase transition, we plot the evolution of the OMM at points $\bm K$ and $\bm K'$ as a function of $\phi^{c}$, as shown in Fig.\ref{Fig44}. The behaviour of the OMM at point $\bm K'$ for $\mathcal{M}_{2}(\bm  K',\phi^{a},\phi^{c})$ and $\mathcal{M}_{1}(\bm  K',\phi^{a},\phi^{c})$ shows that they are monotonic increasing and decreasing functions of $\phi_c$. Also, $\mathcal{M}_{0}(\bm  K,\phi^{a},\phi^{c})$ and $\mathcal{M}_{1}(\bm  K,\phi^{a},\phi^{c})$ are monotonically decreasing and increasing functions at the $\bm K$ point. Conversely, the OMM for $\mathcal{M}_{2}(\bm  K',\phi^{a},\phi^{c})$ and $\mathcal{M}_{1}(\bm  K',\phi^{a},\phi^{c})$ at point $\bm K'$ changes sign arbitrarily, going from positive to negative through $\phi^c = \pi/6$. Furthermore, we observe an inversion of the sign of the OMM at point $\bm K$ for $\mathcal{M}_{0}(\bm  K,\phi^{a},\phi^{c})$ and $\mathcal{M}_{1}(\bm  K,\phi^{a},\phi^{c})$ through $\phi^c = 5\pi/6$. We conclude that the OMM exhibits significant divergence at $\phi^c = \pi/6$ and $\phi^c = 5\pi/6$, accompanied by distinct sign changes, which is a clear indication of a topological phase transition.
\section{Chern phase diagram}\label{secV}
In this section, we aim to detect the topological phase transitions that were previously identified in terms of the OMM and the Berry curvature. We explore these transitions using the Chern number, which is a basic topological invariant used to characterise the topological phases of matter, especially in Chern insulators. The Chern number is an integer that quantifies the topology of electronic bands within the first Brillouin zone. If it is non-zero, this implies the existence of conductor edge states, which are protected from external perturbations \cite{ReffChern1}. This phenomenon is a strong characteristic of topological phases. In mathematical terms, the Chern number is defined as the integral of the Berry curvature over the first Brillouin zone, corresponding to the $\nu$-th band \cite{ReffChern4A,ReffChern4}. It is also known as the TKNN index \cite{In11,ReffChern3} and is written as follows:
\begin{equation}
C_{\nu}=\dfrac{1}{2\pi}\int_{FBZ}\Omega_{\nu}(\bm  k,\phi_{a},\phi_{c})d\bm k.
\end{equation}
NNN hopping is essential for breaking time-reversal symmetry without a magnetic field, which enables us to study the topological evolution of the $\alpha-T_{3}$ lattice. In particular, we consider a phase configuration in which the NNN hops differ, i.e., in which the $\phi^a$ and $\phi^c$ phases are distinct. This difference results in three topologically non-trivial bands: the flat band ($\nu = 1$), the conduction band ($\nu = 2$), and the valence band ($\nu = 0$). To analyse the topological nature of the system, we plot the Chern number as a function of the phases $\phi^{a}$ and $\phi^{c}$ in two distinct cases. In the \textbf{first case}, we fix $\alpha = 1$ and set $\phi^a = \phi^c = \phi$, and then plot the Chern number $C_{\nu}$ as a function of $\phi$ over the interval $[0, \pi]$. In the \textbf{second case}, we set $\phi^{a}$ to $\pi/2$ and plot $C_{\nu}$ as a function of $\phi^{c}$ over the interval $[0, \pi]$.\newline
\begin{figure*}[!t]
	\centering
	\begin{tikzpicture}
	\node[anchor=center] at (1,0) {\includegraphics[width=0.4\linewidth]{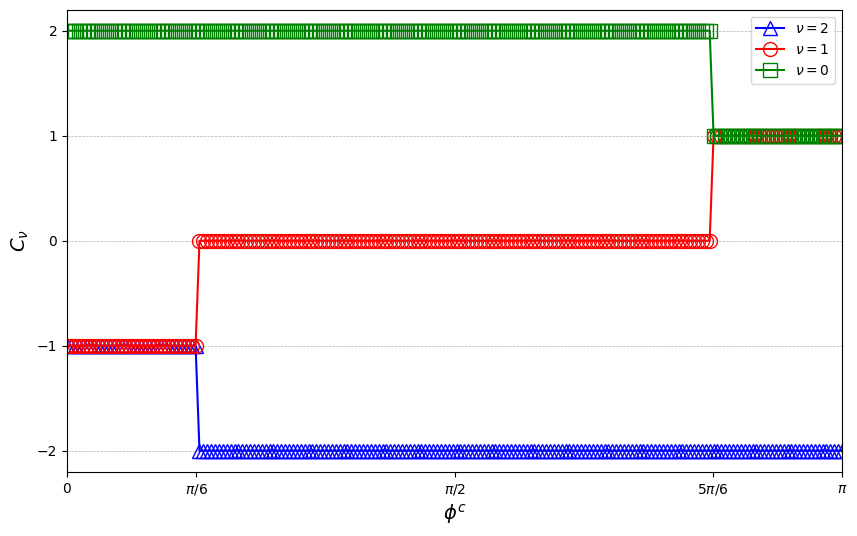}};
	\node at (-0.9,1.7) { (a) \quad $\phi^a=\dfrac{\pi}{2}$ };  % <- Symbole au centre
	\end{tikzpicture}
	\begin{tikzpicture}
	\node[anchor=center] at (1,0) {\includegraphics[width=0.4\linewidth]{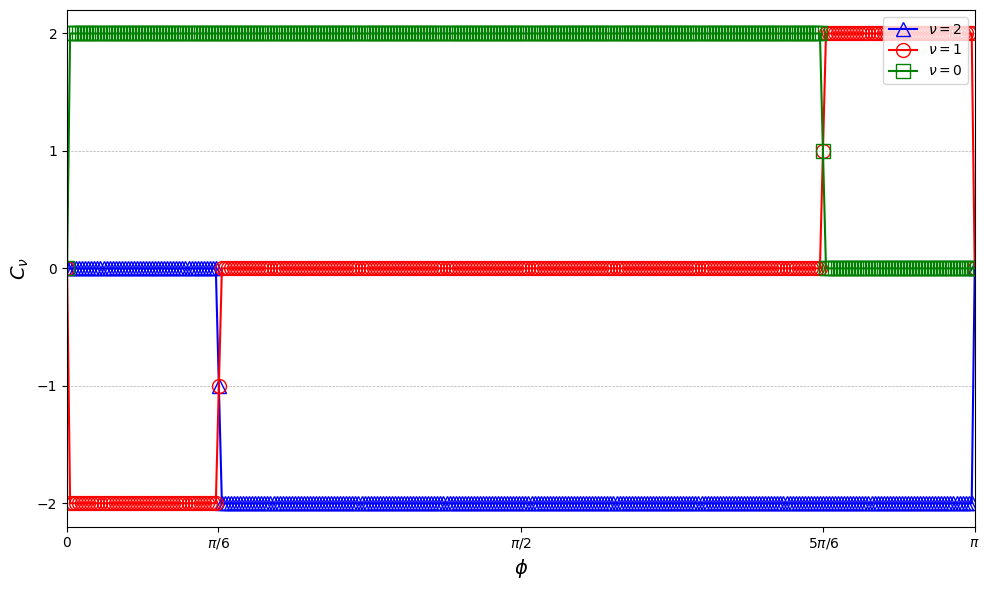}};
	\node at (-0.7,1.7) { (b) \quad $\phi^a=\phi^c=\phi$ };  % <- Symbole au centre
	\end{tikzpicture}
	\caption{(a) Variation of the Chern number as a function of $\phi^{c}$, with $\phi^{a}$ fixed at $\pi/2$. (b) Variation of the Chern number as a function of $\phi = \phi^{c} = \phi^{a}$. The variation of the Chern number corresponding to the valence band ($\nu=0$), the flat band ($\nu=1$), and the conduction band ($\nu=2$) is represented by green, red, and blue, respectively, as shown in subfigures (a) and (b). The other parameters are set to $\alpha=1$ and $t_{a} = t_{c} = 0.1t$.}
	\label{Fig55}
\end{figure*}
We begin by analyzing the \textbf{first case}. The Chern number evolves continuously over the interval $[0, \pi]$, as illustrated in Fig.\ref{Fig55}(b). We observe that the system becomes topological as soon as the Chern number becomes non-zero. This continuous variation in $\phi$ reveals a topological phase transition, which is marked by a crossing of the flat band and the conduction band at the Dirac points $\bm K$ and $\bm K'$ for $\phi = \pi/6$, and a crossing of the flat band and the valence band at the same points for $\phi = 5\pi/6$.\newline
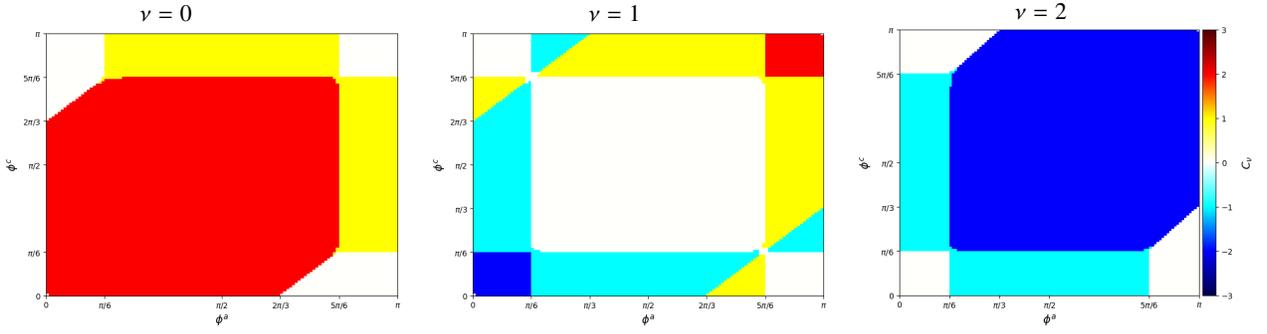
\begin{figure*}[!t]
	\centering
	\begin{tikzpicture}
	\node[anchor=center] at (1,0) {\includegraphics[width=0.3\linewidth]{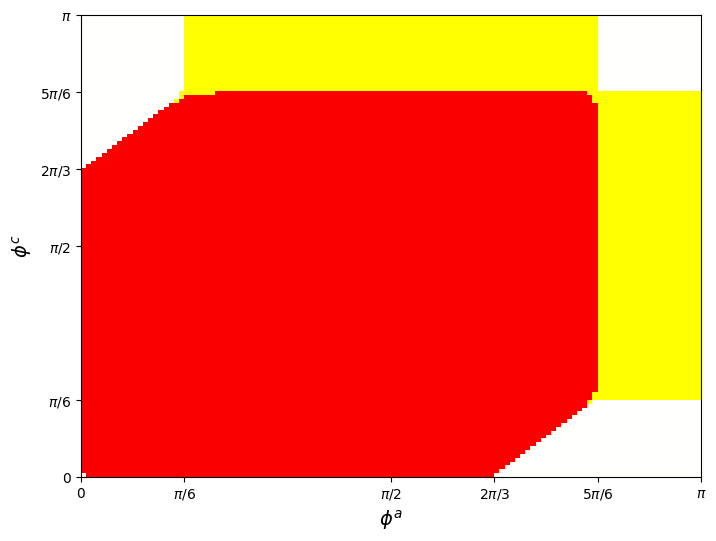}};
	\node at (0.5,2.2) { $\nu=0$ };  % <- Symbole au centre
	\end{tikzpicture}
	\begin{tikzpicture}
	\node[anchor=center] at (0.7,0) {\includegraphics[width=0.3\linewidth]{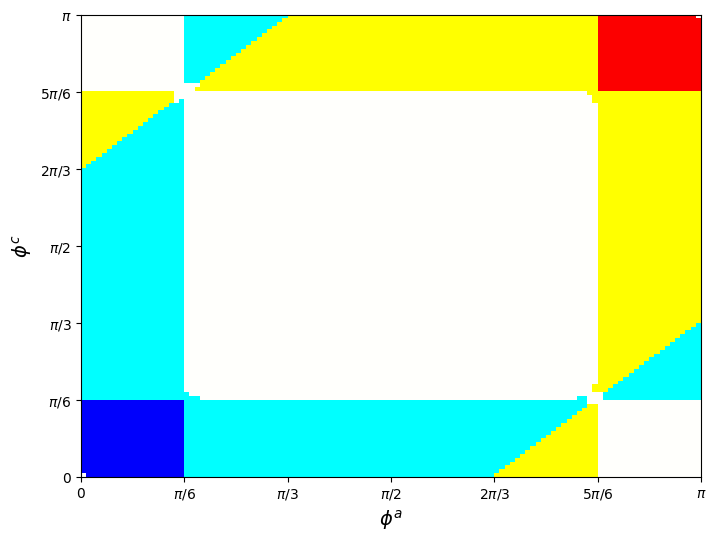}};
	\node at (0.5,2.2) { $\nu=1$ };  % <- Symbole au centre
	\end{tikzpicture}
	\begin{tikzpicture}
	\node[anchor=center] at (0.7,0) {\includegraphics[width=0.3\linewidth]{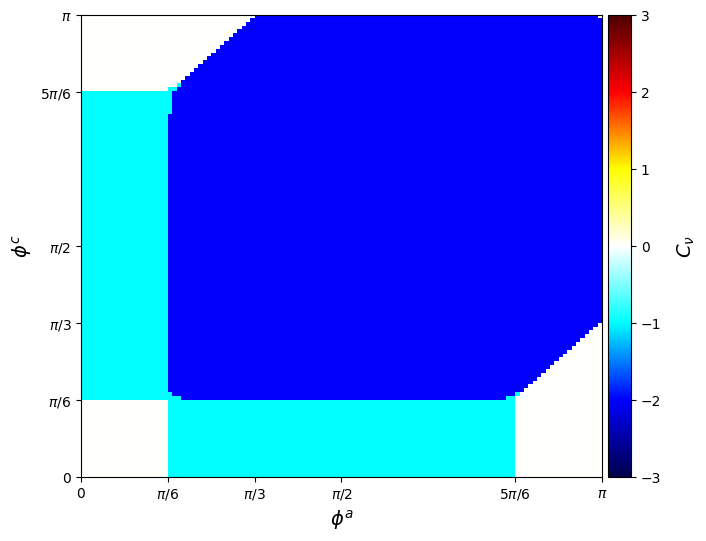}};
	\node at (0.5,2.2) { $\nu=2$ };  % <- Symbole au centre
	\end{tikzpicture}
	\caption{The Chern number phase diagram is shown in flux space ($\phi^c$ and $\phi^a$), which corresponds to the valence band ($\nu=0$), the flat band ($\nu=1$), and the conduction band ($\nu=2$). The other parameters are set to $\alpha = 1$ and $t_{c} = t_{a} = 0.1t$.}
	\label{Fig555}
\end{figure*}
\par When $\phi$ belongs to the interval $\left[ 0, \pi/6\right[$, the Chern number associated with the flat (or valence) band is equal to -2 (or 2, respectively), indicating a non-trivial topological nature. The conduction band, on the other hand, has a zero Chern number and is therefore topologically trivial. In the $\left] \pi/6, 5\pi/6\right[$ region, the system undergoes a topological transition at $\phi = \pi/6$. At this value, the Chern number of the flat band becomes zero, rendering it topologically trivial. Conversely, the conduction band becomes topologically non-trivial, with a Chern number  -2, while the valence band remains unchanged as it does not participate in the gap closure that occurs between the flat band and the conduction band at points $\bm K'$ and $\bm K$, where the flat band becomes dispersive. Another topological transition is observed at $\phi = 5\pi/6$ in the interval $\left] 5\pi/6, \pi\right[$. At this point, the initially topologically trivial flat band ($C_{1} = 0$) becomes non-trivial with a Chern number  2, while the previously non-trivial valence band ($C_{0} = 2$) becomes trivial with a Chern number  zero. The conduction band remains unchanged as it does not participate in the band crossover at the $\bm K'$ and $\bm K$ points, where the flat band connects to the valence band and also becomes dispersive.\newline
 In the \textbf{second case}, we set $\phi^a=\pi/2$ and vary $\phi^c$, thereby breaking the inversion symmetry. As illustrated in Fig.\ref{Fig55}(a), we find that the distribution of Chern numbers differs in this case compared to the first scenario. In the first region $\left[ 0, \pi/6\right[$ , we observe that the flat and conduction bands both have a Chern number -1, indicating a non-trivial topological nature, whereas the valence band is topologically non-trivial with a Chern number 2. This behaviour differs from that in the previous case (when $\phi^a =\phi^c$), where the conduction band was topologically trivial. A phase transition is identified at $\phi^c =\pi/6$. In the $\left] \pi/6, 5\pi/6\right[$ region, the system undergoes a topological change: the flat band becomes topologically trivial (Chern number zero), while the conduction band remains topologically non-trivial, with a Chern number evolving from -1 to -2; the valence band, however, remains topologically non-trivial as it does not participate in gap closure. Gap closure occurs between the flat band and the conduction band at the Dirac point $\bm K'$, where the flat band becomes dispersive.\newline
A second phase transition occurs at $\phi^c=5\pi/6$ in the region $\left] 5\pi/6, \pi\right[$ . The system changes its topological structure again: the valence band, which initially had a Chern number 2, now has a Chern number 1, while the flat band changes from being topologically trivial $(C_{1} = 0)$ to being topologically non-trivial $(C_{1} = 1)$. This behaviour contrasts with that observed in the first case $(\phi^a = \phi^c)$, where the valence band became trivial after the transition. The key difference here is the asymmetry of $\phi^a \neq \phi^c$, which keeps the system in a topologically non-trivial regime for all bands in this region. This asymmetry breaks the inversion symmetry. Meanwhile, the conduction band retains its topologically non-trivial nature throughout, and gap closure only occurs between the flat band and the valence band at the Dirac $\bm K$ point at $\phi^c = 5\pi/6$, as previously observed (see Fig.\ref{Fig0}(b)). The $\alpha-T_{3}$ lattice exhibits intriguing topological properties when its time and inversion symmetries are simultaneously broken, a phenomenon influenced by the $\phi^{c}$ and $\phi^{a}$ fluxes. Variation of these two parameters together leads to the emergence of multiple topological phases. To explore how the system's topological properties evolve, we plot the Chern number in a phase diagram as a function of the $\phi^{c}$ and $\phi^{a}$ fluxes, each varying within the range $\left[ 0, \pi\right]$, as illustrated in Fig.\ref{Fig555}. This approach enables us to identify potential additional topological transitions and synthesise the two previously analysed cases —  namely, $\phi^{c} =\phi^{a}$ and $\phi^{c} \neq\phi^{a}$ — within a single framework, using the phase diagram corresponding to the system's three bands. We observe that the phase transition identified earlier, where $\phi^{a}$ is fixed at $\pi/2$ and $\phi^{c}$ varies, occurs at the critical values of $\phi^{c} = \pi/6$ and $\phi^{c} = 5\pi/6$. Conversely, phase transitions appear when $\phi^{c} = \pi/2$ and $\phi^{a}$ varies, at $\phi^{a} =\pi/6$ and $\phi^{a} =5\pi/6$.\newline
When $\phi^{c} \neq\phi^{a}$, new topological transitions appear that are not present in symmetrical cases. In particular, valence-band transitions are identified in the regions  $\phi^{c}\cup\phi^{a}=\left[ 2\pi/3, 5\pi/6\right]\cup \left[ 0, \pi/6\right]$, and conversely in the regions $ \left[ 0, \pi/6\right] \cup \left[ 2\pi/3, 5\pi/6\right] $. A novel topological transition is also observed in the flat band, localised to the regions  $\left[ 2\pi/3, \pi\right] \cup \left[ 0, \pi/3\right]$, and another transition is found in the regions $\left[ 0, \pi/3\right]\cup\left[ 2\pi/3, \pi\right]$. Concerning the conduction band, the phase transitions observed in the previous two cases are preserved globally. However, new transitions also appear, particularly in the regions $\left[ 5\pi/6, \pi\right]\cup\left[ \pi/6, \pi/3\right]$, as well as in the reciprocal regions   $ \left[ \pi/6, \pi/3\right]\cup\left[5\pi/6, \pi\right]$.\newline
\textbf{Note}: The flat band is strongly influenced by variations in the $\phi^{a} $ and $\phi^{c}$ fluxes, which can render it topologically non-trivial with a Chern number up to  $\pm2$. This behaviour has not been observed in other studies, except in those incorporating a Semenoff mass term alongside anisotropy-induced deformation in the jump energy. In this case, the Chern number is equal to  $\pm2$. However, in the absence of deformation and when the Semenoff mass is zero, the flat band remains topologically trivial with a zero Chern number \cite{ReffChern4}. Conversely, in a bilayer $\alpha-T_{3}$ lattice with cyclic stacking for $\alpha = 1$ and $\phi=\pi/2$, the flat band can acquire a Chern number $\pm2$ \cite{ReffChern5}. These observations suggest that considering a controlled flow of $\phi^{a}  \neq \phi^{c} $ or $\phi^{a}  = \phi^{c} $ is essential, as this allows the flat band to achieve a high Chern number and reveal a rich, non-trivial topological phase. When point symmetry of charge conjugation is broken ($\phi^{a}  \neq \phi^{c} $), we observe the emergence of a transition associated with the closure of the gap at a single Dirac point, either at $\bm K$ or at $\bm K'$ (e.g, $\phi^{a}  =\pi/2$,  $\phi^{c}  =\pi/6$, et  $\phi^{c}  =2\pi/3$) . This behaviour differs from the case where charge conjugation symmetry is preserved, for which the gap closes partially at both points $\bm K$ and  $\bm K'$. To better understand this phenomenon, let us consider, for example, $\phi^{a}  =\pi/2$ with $\phi^{c}  =\pi/6$, as well as the case  $\phi^{c}  =2\pi/3$. In the phase diagram (Fig.\ref{Fig555}), we observe a single topological transition for $\phi^{c}  =\pi/6$ in the conduction band ($\nu=2$) and the flat band ($\nu=1$), while no transition is observed in the valence band ($\nu=0$). On the other hand, for $\phi^{c}  =2\pi/3$, no transition is observed.
\section{ANOMALOUS HALL CONDUCTIVITY}\label{secVI}
\begin{figure}[H]
 	\centering
 	\includegraphics[width=1\linewidth]{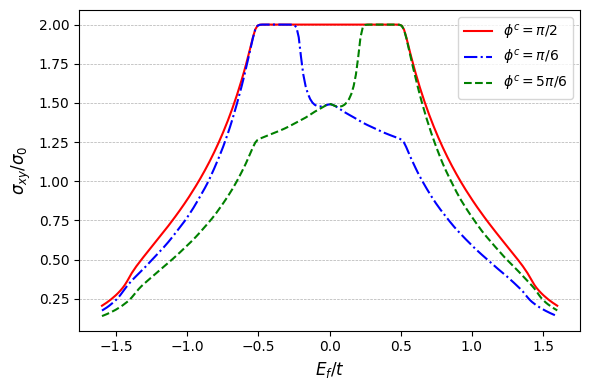}
 	\caption{The plot shows the Hall conductivity as a function of the Fermi energy for three values of $\phi^c$: $\phi^c=\pi/2$ (the standard case, where $\phi^c=\phi^a$), $\phi^c=\pi/6$ (the first transition, where $\phi^c\neq\phi^a$), and $\phi^c=5\pi/6$ (the second transition), represented in red, blue, and green, respectively. The other parameters are set to $\phi^a=\pi/2$, $\alpha=1$, and $t_c = t_a = 0.1t$.}
 	\label{Fig6}
 \end{figure}

In this section, we investigate the anomalous Hall conductivity of the $\alpha-T_{3}$ lattice. This phenomenon occurs when a transverse electric field is generated in a ferromagnetic material in response to an electric current, despite there being no external magnetic field present. This conductivity is intrinsically related to time-reversal symmetry breaking, caused here by the NNN interaction. This provides insight into the topological properties of the $\alpha-T_{3}$ lattice, particularly when the $\phi^{a}$ and $\phi^{c}$ fluxes are controlled. Anomalous Hall conductivity arises from an intrinsic mechanism directly related to the electronic band structure and Berry curvature. The latter induces anomalous electron velocities, thereby contributing to anomalous Hall conductivity \cite{ReffAHCa1}. We calculate conductivity by integrating the Berry curvature across all occupied electronic states within the first Brillouin zone. This is achieved by combining the individual contributions of each $\alpha-T_{3}$ lattice band  \cite{ReffBerry1,ReffAHC1}.
\begin{eqnarray}
\sigma_{xy}\left(E_f,\phi^a,\phi^c\right)=\frac{\sigma_{0}}{2\pi}\sum_\nu \int_{BZ} d^{2}\bm k\, \Omega_{\nu}(\bm  k,\phi_{a},\phi_{c})f_{\nu}(E_{f},\bm k).
\end{eqnarray}
The notation $\sigma_{0} =e^2/h$ represents the elementary quantum conductivity. The function 
$f_{\nu}(E_{f},\bm k)$ is the Fermi-Dirac distribution $\left[e^{(E_{\nu}(\vect k)-E_f)/k_{B}T} +1\right]^{-1}$ with  fixed temperature at $T=100 K$. $k_B$ denotes the Boltzmann constant, $E_F$ the Fermi energy, $E_\nu(\bm k)$ the eigenvalue of the $\nu-th$ band obtained from equation (\ref{EQ12}), and $\Omega_{\nu}(\bm  k,\phi^{a},\phi^{c})$ the Berry curvature associated with the $\nu-th$ band from equation (\ref{EQ16}). We study the anomalous Hall conductivity by setting $\alpha = 1$, keeping $\phi^{a} =\pi/2$, and varying $\phi^{c}$ for three distinct values: $\phi^{c}=\pi/6, \pi/2$, and $5\pi/6$.
 The presence of a non-zero Berry curvature results in a contribution to the anomalous conductivity for all occupied electronic states. If the Fermi level lies within a band gap, the Fermi–Dirac function $f(E_f)$ approaches 1 at low temperatures, meaning that only states below $E_f$ participate. In this case, the integral of the Berry curvature over the Brillouin zone is used to calculate the Chern number, resulting in a quantized conductivity of $\sigma_{xy}\left( E_f,\phi^a,\phi^c\right) =C_{\nu}\sigma_{0}$, where $C_{\nu}$ is the Chern number of the occupied bands. Setting $\phi^a =\phi^c= \pi/2$ and $\alpha=1$ causes the conductivity to reach a quantized value of $2\sigma_{0}$, provided the Fermi level ($E_f$) remains within the width of the band gap (as shown in Fi.\ref{Fig6}). In this case, the flat band does not contribute to the Berry curvature ($\Omega_1 =0$), and the width of the plateau corresponds to that of the band between the valence and conduction bands. The obtained conductivity value is the same as that found in references \cite{IIn2,ReffChern4A,ReffChern5}. When the Fermi level $E_f$ exceeds the band gap, the conductivity $\sigma_{xy}\left( E_f,\phi^a,\phi^c\right)$ begins to decrease because the integration is performed on unoccupied electronic states beyond the edge of the gap.
 \noindent When $E_{f}$ exceeds this gap, $\sigma_{xy}\left( E_f,\phi^a,\phi^c\right)$ starts to decrease, the integral is then performed on occupied electronic states.
When $\phi^a\neq\phi^c$, with $\phi^a=\pi/2$  and $\phi^c=\pi/6$ and $5\pi/6$, the flat band becomes dispersive. This leads to a change in Berry curvature and an effective reduction in the gap width. This reduction in the gap decreases the overall contribution of the occupied electronic states, as illustrated in Fig.\ref{Fig6} for $\phi^c=\pi/6$. In this regime, the width of the Hall conductivity plateau decreases proportionally to the gap between the valence band and the dispersive flat band. The Hall conductivity then reaches a quantized value, $\sigma_{xy}\left( E_f,\phi^a,\phi^c\right) = 2\sigma_{0}$. When the Fermi level $E_F$ leaves this zone, $\sigma_{xy}\left( E_f,\phi^a,\phi^c\right)$ decreases and adopts a non-quantised value, $\sigma_{xy}\left( E_f,\phi^a,\phi^c\right)=1.50  \sigma_{0}$. It then takes the form of a linear inclined plateau, gradually decreasing from $\sigma_{xy}\left( E_f,\phi^a,\phi^c\right)=1.50 \sigma_{0}$ to $\sigma_{xy}\left( E_f,\phi^a,\phi^c\right)=1.25 \sigma_{0}$. This non-quantized value arises from the fact that the flat band, becoming dispersive, generates a non-zero Berry curvature. Integration of this curvature within the first Brillouin zone relates solely to partially occupied states since the Fermi level crosses incompletely filled bands. This mechanism explains the intermediate value of the Hall conductivity. Similar behaviour is observed for $\phi^c=5\pi/6$, with a further reduction in plateau width when the Fermi level lies within the gap between the dispersive flat band and the conduction band. The Hall conductivity first reaches its quantized value of 2$\sigma_{0}$, and then decreases to form an inclined plateau ranging from 1.50$\sigma_{0}$ to 1.25$\sigma_{0}$. In both cases, this decrease in conductivity is due to the contribution of the Berry curvature associated with the dispersive flat band. Unquantified values appear precisely when the Fermi level crosses partially occupied bands. The obtained Hall conductivity is unquantised, with values of $\sigma_{xy} =1.5\sigma_{0}$ when the Fermi level is close to $F_f\approx0$ and  $\sigma_{xy} =1.25\sigma_{0}$ for $E_f\approx\pm 0.5t$. This behaviour can be explained by the fact that the flat band becomes dispersive for $\phi^c=\pi/6$ and $\phi^c=5\pi/6$. This dispersion induces a non-zero, non-uniform Berry curvature in the $k_x-k_y$ plane of reciprocal space. Integrating the Berry curvature over the occupied states then generates partial contributions from several topological bands located near the Fermi level. As these bands are only partially occupied, the integral is not taken over the entire Brillouin zone, and the resulting conductivity does not correspond to an integer topological invariant. This behaviour stems directly from the dispersive nature of the flat band, combined with the absence of a sufficiently large energy gap to topologically isolate the individual contributions of each band. Consequently, the Hall conductivity is not strictly quantized. A symmetry relation is also observed: the conductivity between the points $\phi^c=\pi/6$ and $\phi^c=5\pi/6$ is equal, i.e. $\sigma_{xy}\left( E_f,\pi/2,5\pi/6\right) =\sigma_{xy}\left( -E_f,\pi/2,\pi/6\right) $. The property is a consequence of the point-charge conjugation symmetry of the system. The anomalous Hall conductivity in the dice lattice is highly sensitive to the controlled flux between sites. This flux breaks the inversion symmetry, directly influencing the topology of the bands and the properties of quantum transport.
\section{ Conclusion.}\label{secVII}
In this research work, we examined how the Haldane model can be extended to the $\alpha-T_3$ lattice. This was achieved by introducing two modifying fluxes, $\phi^{a}$ and $\phi^{c}$, between the sites of sub-lattice A and C, respectively. We demonstrated that this coupling results in the breaking of time reversal symmetry for $\alpha=1$. When $\phi^{a} = \phi^{c}$ , however, the inversion, charge conjugation and $C_3$ rotation symmetries are preserved. In this case, our results are consistent with those of previous studies \cite{IIn2,ReffChern4A,ReffChern5}. Conversely, when $\phi^{a} \neq \phi^{c}$, inversion symmetry is broken. However, charge conjugation symmetry may remain as a point symmetry in certain special cases. This behaviour is not present in either the Haldane standard model or the aforementioned previous work, which highlights the originality of our approach.  We have also derived an exact analytical expression for the quasi-energy. When $\phi^{a} = \phi^{c} = \pi/2$, lifting the degeneracy produces equally separated energy bands. Conversely, when $\phi_{a} \neq \phi_{c}$, the gap closes, signalling a transition back to a semimetallic phase marked by the formation of low-energy Dirac cones at points $\bm K$ and $\bm K'$. Thus, the gap’s behaviour precisely signifies a topological phase transition, which is clearly illustrated by analysing the Berry curvature and the orbital magnetic moment. 
We observed several topological phase transitions in the phase diagram. For values of the parameter $\phi^{a}$  equal to $\pi/2$ and different from $\phi^{c}$, specifically in the region where $\phi^{c} \in [0, \pi/6[$, the Chern numbers $(C_{2}, C_1)$ take the values $(-1,-1)$, unlike the scenario where both parameters are equal ($\phi^{a} = \phi^{c} \in[0, \pi/6[$), which yields $(C_{2}, C_{1}) = (-2, 0)$. This indicates a change in the topological nature of the system. Similarly, in the region $\phi^{c} \in]5\pi/6, \pi]$, the Chern numbers $(C_{1}, C^{0})$ become (1,1) for $\phi^{a}  = \pi/2 \neq \phi_{c} $, while for $\phi^{a} = \phi^{c} \in]5\pi/6, \pi]$, they are (2,0). This difference highlights a transition between a trivial insulating phase and a non-trivial phase, a result absent from previous studies \cite{IIn2,ReffChern4A,ReffChern5}.
Finally, analysis of the Hall conductivity reveals a quantized plateau at $\phi^{a} =\phi^{c} =\pi/2$, whose width is proportional to the gap and whose value is 2$\sigma_{0}$. When $\phi_{c}$ deviates from this condition, the plateau's width and the conductivity's value decrease. The plateau then loses its quantization and becomes inclined and quasi-linear. Similar behaviour is observed for $\phi^{c}$, in accordance with the symmetry relation $\sigma_{xy}(E_{f},\phi^{a},\phi^{c}_{2}) = \sigma_{xy}(-E_{f},\phi^{a},\phi^{c}_{1})$.\\
These results emphasise the variety of topological phases that can arise from changes in flow in the extended $\alpha-T_3$ model. This paves the way for new possibilities in topological control of systems inspired by the Haldane model. Flux tuning and the subsequent opening and closing of the gap at the Dirac points is an effective methods of engineering topological phase transitions. This enables transitions between the semimetallic and topological insulator phases, opening up the possibility of developing low-dissipation, reshapeable electronic and optical devices such as low-power topological transistors \cite{R1r1}, as well as applications in spintronics and robust quantum transport \cite{R1r2,R1r3}.
\section*{Experimental proposal}
\par This highlights the importance of the experimental framework. In fact, experimental verification of our model hinges on the ability to control the coupling conditions that cause symmetry breaking and the associated phase transitions independently. We propose two complementary experimental platforms through which the correlations illustrated in Fig.\ref{Fig0} can be realised, modified and characterised: inhomogeneous structures in the solid state and ultracold atomic gases trapped in optical lattices.
\subsection*{ Solid-state implementation: Van der Waals heterostructures and Floquet engineering}
\noindent In the field of solid-state physics, the proposed model can be realised using van der Waals heterostructures. The experimental realisation of this model proceeds by simultaneously breaking inversion and time-reversal symmetries through distinct mechanisms, providing the basis for the implementation outlined. Firstly, inversion symmetry breaking ($\mathcal{I}$) is achieved through substrate engineering. This is done by depositing the dice lattice ($\alpha = 1$) on hexagonal boron nitride (h-BN). The structural similarity between h-BN and the dice lattice allows for precise alignment of the B and C sublattices with the boron and nitrogen atoms. This alignment induces distinct local electrostatic potentials. Such an on-site energy imbalance generates a Semenoff-type mass term. The effectiveness of this technique has already been demonstrated experimentally for graphene \cite{R1r4}. At the same time, time reversal symmetry ($\mathcal{T}$) is broken dynamically by Floquet engineering. This approach involves applying a circularly polarised laser field in a non-resonant regime (i.e. the high-frequency limit). This generates purely imaginary next-nearest-neighbour (NNN) hopping terms analogous to the Haldane term through light-matter interaction. Physically, the system can be described as a light-dressed state involving virtual photon emission and absorption processes. The spectroscopic signature of these Floquet-Bloch states has also been observed experimentally on the surface of the topological insulator $Bi_2 Se_3$ using time- and angle-resolved photoemission spectroscopy (TR-ARPES) \cite{R1r5}.  When this method is applied to break this symmetry of inversion and reversal of time simultaneously and control the parameters such as H-BN subtraction and the amplitude of circular polarization and frequency, we can achieve the results of this probable model.\newline 
\subsection*{Implementation in an optical lattice (cold atoms)}
\noindent As an alternative approach, we propose to realize this model using an optical lattice platform, following the scheme introduced by Bercioux et al.\cite{ In49}, implemented with ultracold fermionic atoms. In this configuration, the lattice is generated by three pairs of counter-propagating laser beams. From a geometrical perspective, the interference of these beams divides the plane into six sectors of $60^{\circ}$. All lasers share the same wavelength $\lambda=3/2\lambda_{0}$, where $\lambda_{0}$ denotes the lattice constant. The beams are linearly polarized in the $xy$-plane: one pair is polarized along the y-axis, while the other two pairs have their polarization vectors rotated by $120^{\circ}$ around the z-axis.
The resulting interference pattern creates a crystalline structure whose unit cell contains three inequivalent sites: a hub site C with coordination number 6, and two rim sites A and B, each with coordination number 3. This configuration naturally realizes massless Dirac–Weyl fermions in the low-energy spectrum.
To break inversion symmetry in a controlled manner, we introduce an additional pair of laser beams rotated around the z-axis and slightly detuned in frequency. This generates a periodic optical potential that induces an energy offset between the sublattices, thereby opening a gap at the Dirac points. Such a method was proposed by L. Tarruell et al. \cite{R1r7}. Simultaneously, time-reversal symmetry can be broken by applying off-resonant circularly polarized light. Within the framework of Floquet engineering, this induces an effective next-nearest-neighbor (NNN) hopping term of the standard Haldane type \cite{R1r8}. Combining these techniques allows us to realise the model and study its phase transitions by adjusting the modified flux, which simultaneously breaks inversion and time reversal symmetries.

\end{document}